\documentclass{jpsj2}

\def\runtitle{Effects of Electron Correlation on the Transport through 
a Quantum Dot Superlattice}
\def\runauthor{Yoshihide \textsc{Tanaka} and Akira \textsc{Oguri} }

\title{
Effects of Electron Correlation on the Transport through \\ 
a Quantum Dot Superlattice
}

\author{
Yoshihide \textsc{Tanaka}\thanks{E-mail tanakay@sci.osaka-cu.ac.jp} and 
Akira \textsc{Oguri}
}

\inst{
Department of Material Science, 
Osaka City University, Sumiyoshi-ku, Osaka 558-8585, Japan
}

\recdate{March 14, 2003}

\abst{
We study correlation effects on 
the transport through a quantum dot superlattice 
using a two-dimensional Hubbard model 
connected to two noninteracting leads. 
To calculate the zero-temperature conductance away from half-filling,
we have used the non-magnetic solution of the Hartree-Fock approximation 
for the unperturbed Green's function, and then take into account the electron
correlation beyond the mean field theory through 
the second-order self-energy corrections with respect 
to the residual interaction.
When the value of the onsite potential $\epsilon_d$ or the repulsion $U$ is 
changed, the conductance shows a number of peaks corresponding to 
the resonance states passing through the Fermi energy. 
The behavior of the resonance states for correlated electrons 
can be investigated using the eigenvalue of
the effective Hamiltonian for a quasi-particle 
of a local Fermi liquid. 
We calculate the eigenvalue using the second-order self-energy, 
and show explicitly that it can be compared with the conductance peak.  
}

\kword{
conductance, electron correlation, Hubbard model, 
two-dimension, perturbation expansion, electron-hole asymmetry}

\begin{document}
\sloppy
\maketitle

\section{Introduction}

Systems of nanometer scale and artificial structures, 
such as quantum dots and quantum wires, 
have been a subject of current interest.
In these systems the Coulomb interaction seems to play an important role 
on the transport properties. 
For instance, the Kondo effect
in quantum dots has been studied intensively~\cite{NL,GR,Kawabata,MWL1-2,HDW2} 
and  the experiments have been shown to be in qualitative agreement 
with theoretical predictions.~\cite{Goldharber-Gordon,Kouwenhoven} 
The artificial molecule which can be realized 
by arranging two or more quantum dots 
is another example.~\cite{oosterkamp,tokura} 
Also, an interesting quantum dot superlattice 
showing the ferromagnetism or \textit{d}-wave superconductivity 
has been designed theoretically.~\cite{tamura1-4,kimura} 
The proposed superlattice consists of a two-dimensional network 
of quantum wires, and is described by a Hubbard model that 
is derived based on the local density approximation.

In theoretical studies of 
the transport properties of strongly correlated electrons, 
accurate numerical approaches such as 
the numerical renormalization group~\cite{Izumida1-3,Izumida4-5} and 
quantum Monte Carlo methods~\cite{Sakai,ao6} have been used
successfully. However, the analytic methods, 
which are complementary to the numerical ones, 
are also required to clarify the underlying physics comprehensively.  
The perturbation theory in the Coulomb interaction 
is one of such approaches. 
For instance in the Kondo problem,\cite{Hewson} 
the perturbation expansion with respect to $U$ has shown to work 
quite well to describe the low-energy 
Fermi-liquid behavior.\cite{YamadaYosida,ZlaticHorvatic} 
The perturbation theory can also describe properly the low-voltage behavior of 
the out-of-equilibrium Kondo system under 
finite bias voltages.\cite{HDW2,ao11-ao12} 
Furthermore, based on the analytic 
properties of the vertex function,
the transport equation for the Fermi liquid derived by 
\'Eliashberg~\cite{Eliashberg} 
has been shown to have a link to 
the Landauer formula for the conductance of 
the correlated electrons.~\cite{ao10}

\begin{figure}[tb]
\begin{center}
\leavevmode
\includegraphics[width=0.7\linewidth, clip, 
trim = 3cm 10cm 3cm 10cm]{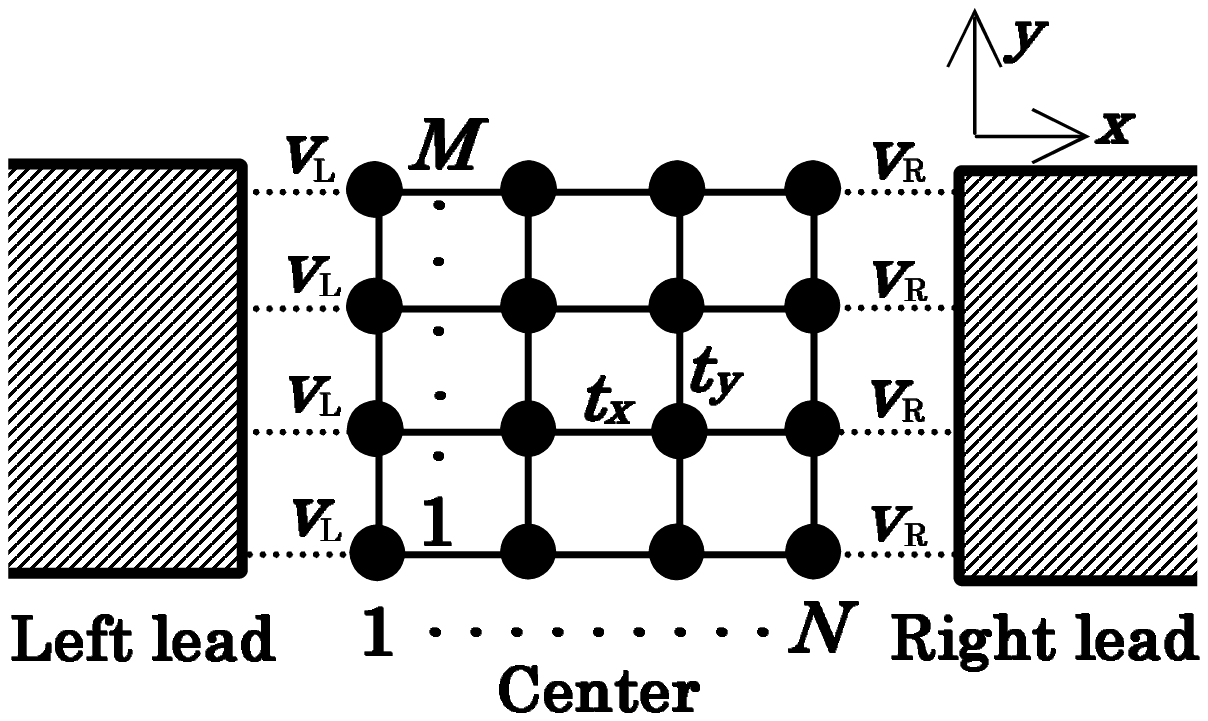}
\caption{Schematic picture of the model:
($\bullet$) interacting sites.} 
\label{fig:fig01}
\end{center}
\end{figure}

We have been studying the transport properties of a quantum dot superlattice 
based on a Hubbard model coupled to two noninteracting leads.~\cite{YT} 
A schematic picture of the model is illustrated in Fig.\ \ref{fig:fig01}: 
the system consists of a Hubbard model of $N_{\rm C}$ ($=N \times M$) sites 
and two semi-infinite noninteracting leads of $M$ channels at left and right, 
where $N$ and $M$ are the size in the $x$- and $y$-direction, respectively. 
We have considered the square lattice for simplicity. 
At this moment this model itself might have no direct correspondence 
to experiments. Nevertheless, it seems to contain 
essential features of two-dimensional network of quantum dots 
or atomic systems, and we believe that 
it could be realized some day in near future.
In this kind of the system that coupled to the noninteracting leads,
the contribution of the vertex corrections on the conductance   
vanishes at zero temperature.~\cite{ao7-ao9} 
Thus, at $T=0$, the conductance can be written in terms of 
the single-particle Green's function at the Fermi energy $\omega=0$. 
In previous work we have calculated the conductance at half-filling,  
where the average number of the electron per site is one,
using the order $U^2$ self-energy.
The conductance shows some peaks as a function of $U$   
when the hopping matrix elements are anisotropic.
We found at half-filling that the position of the resonance peaks 
can be analyzed through the eigenvalues of the effective 
Hamiltonian for free quasi-particles.
However, in actual quantum dots  
the half-filled situation is expected only for  
a special value of the gate voltage, which can be related to 
the onsite energy $\epsilon_d$ of the theoretical model.  
Therefore, real systems are usually in away from half-filling, 
so that it should be examined on the same basis.

The purpose of this paper is to study the correlation effects  
on the transport properties away from half-filling. 
In this case, the average number of electrons per site is not unity,  
and the spatial distribution of the charge is not uniform. 
Therefore, as the density functional theory states,\cite{HohenbergKohn}
the charge distribution should be determined accurately 
to obtain the ground-state properties correctly. 
Note that the charge distribution is uniform at half-filling, 
so that we did not have to take any care over the details 
of the distribution in the previous study.
In the present study, we have used a non-magnetic solution of 
the Hartree-Fock approximation 
to determine the unperturbed Green's function, 
and then take into account the correlation effects beyond 
the mean field approximation through 
the second-order self-energy corrections 
with respect to the residual interaction.
Near half-filling, the result of the conductance for even $N_{\rm C}$  
shows a valley-like structure in the $\epsilon_d$ dependence. 
Comparing with the Hartree-Fock results,
we see that the self-energy corrections make the valley deep and wide. 
This is consistent with the tendency towards the development of 
a Mott-Hubbard insulator gap, and we have examined this feature both 
for one-dimensional chain and two dimensional lattice. 
For odd $N_{\rm C}$ the correlation effects 
make the conductance peak, 
which is seen in the $\epsilon_d$ dependence near half-filling, flat. 
This can be regarded as a manifestation of the Kondo effect, 
and we have carried out the calculations 
for odd $N_{\rm C}$ using a one-dimensional chain.

In the finite system connected to reservoirs, 
the conductance shows a peak 
when a resonance state pass through the Fermi energy. 
Especially, away from half-filling the peak structure can be seen 
by changing the value of $\epsilon_d$. 
The behavior of the resonance states 
can also be analyzed in terms of an extended version of 
a local Fermi-liquid theory. 
The presence of the noninteracting leads makes it possible 
to construct an effective Hamiltonian of a bilinear form, 
which reproduces exactly the conductance and 
the total charge displacement 
of the interacting system at $T=0$. 
In this paper we demonstrate 
using the order $U^2$ self-energy corrections that 
the eigenvalue of the effective Hamiltonian can be  
used also away from half-filling for investigating 
the position of the resonance states.

Horvati\'c and Zlati\'c have shown  
for the single impurity Anderson model that
the radius of the convergence of the 
perturbation expansion with respect to $U$ decreases 
rapidly when the system goes away from half-filling.~\cite{horva1} 
The same thing happens also for systems consisting of 
a number of interacting sites, and away from half-filling 
the second-order perturbation theory is applicable 
for relatively small values of $U$ compared to that at half-filling. 
Therefore, in the present work we have concentrated on 
the weak interaction cases, 
and have checked the applicability by comparing the occupation number 
which are obtained with the two different ways; 
{\em i}) integration of the density of states up to the Fermi level, and 
{\em ii}) the Friedel sum rule. 
The results obtained with these two methods agree well for small $U$.

In \S 2, 
we describe the model and formulations for studying
the correlation effects away from half-filling. 
In \S 3, we present the results of 
the conductance and the average number of electrons, 
and discuss the interpretation based on the effective Hamiltonian. 
Summary is given in \S 4.

\section{Model and Formulation for away from Half-filling}
\label{sec:Hubbard2D}

In the system without the translational invariance, 
the spatial distribution of the charge is usually not uniform, 
except for the electron-hole symmetric case 
where the average number of electrons on each site is fixed as one. 
Therefore, away from half-filling,
the charge distribution depends on the parameters of the Hamiltonian,
and an extra care is necessary to construct the unperturbed part
appropriately.
In this section, we describe the outline of the approach: 
the unperturbed part is determined with 
the Hartree-Fock approximation assuming that 
the ground state to be paramagnetic, 
and then the electron correlation is taken into account 
using the perturbation expansion with respect to the residual interaction.

Fig.\ \ref{fig:fig01} shows the schematic picture of the model, 
where the lattice sites (quantum dots) are labeled with the coordinate 
$\mbox{\boldmath $r$} = (i,j)$. 
The interacting region, where the onsite repulsion $U$ is finite, 
consists of $M$ rows in the $y$-direction 
and $N$ columns in the $x$-direction:  
$t_x$ and $t_y$ are the nearest-neighbor hopping matrix elements 
in the $x$- and $y$-directions, respectively. 
At the columns of $x=1$ and $x=N$, the interacting region 
is connected to noninteracting leads with 
the mixing matrix elements $v_\mathrm{L}$ and $v_\mathrm{R}$, 
which are assumed to be uniform in the $y$-direction. 
The total Hamiltonian $\mathcal{H}$ is given by 
\begin{eqnarray}
\mathcal{H}\ 
&\equiv&  \ \mathcal{H}^0 + \mathcal{H}^\mathrm{int}_\mathrm{C}
\label{eq:H}
\;,
\\ 
\mathcal{H}^0 \ 
&\equiv&  \ \mathcal{H}_{\rm L} + \mathcal{H}_\mathrm{R}
     + \mathcal{H}_\mathrm{C}^0 + \mathcal{H}_\mathrm{mix}
\label{eq:H_0}
\;, 
\\
\mathcal{H}_\mathrm{L} &=&  
\sum_{k \sigma} 
\epsilon_{\mathrm{L} k }^{\phantom{0}}
\,a^{\dagger}_{\mathrm{L} k \sigma} 
a^{\phantom{\dagger}}_{\mathrm{L} k \sigma} 
\;,
\qquad 
\mathcal{H}_\mathrm{R} \, = \, 
\sum_{k\sigma} 
\epsilon_{\mathrm{R} k}^{\phantom{0}}
\,a^{\dagger}_{\mathrm{R} k \sigma} 
a^{\phantom{\dagger}}_{\mathrm{R} k \sigma} 
\;, 
\\
\mathcal{H}_\mathrm{C}^{0} 
&=&  \sum_{i=1}^{N-1} \sum_{j=1}^{M} \sum_{\sigma} 
\left[\, 
-t_x\, c^{\dagger}_{i+1,j, \sigma} c^{\phantom{\dagger}}_{i,j, \sigma} 
\ + \ \mathrm{H.c.}\, \right] 
\nonumber
\\
& &  + 
\sum_{i=1}^{N} \sum_{j=1}^{M} \sum_{\sigma} 
\left[\, 
-t_y\, c^{\dagger}_{i,j+1, \sigma} c^{\phantom{\dagger}}_{i,j, \sigma} 
\ + \ \mathrm{H.c.}\, \right] 
\nonumber
\\
& & + 
\sum_{i=1}^{N} \sum_{j=1}^{M}
\, E_{i,j}\,
 (\, n_{i,j, \uparrow}\, + \,n_{i,j, \downarrow}\, )
\label{eq:H_C_0}
\;,
\\
E_{i,j} 
&\equiv& 
\epsilon_d + \frac{\langle n_{i,j} \rangle_0}{2}\, U 
\;,
\label{eq:E_d}
\\
\mathcal{H}_\mathrm{C}^\mathrm{int} 
&=& 
U 
\sum_{i=1}^N 
\sum_{j=1}^M
\left[\, n_{i,j, \uparrow}\, n_{i,j, \downarrow}
- \frac{\langle n_{i,j}\rangle_0}{2}\,  
( n_{i,j, \uparrow} + n_{i,j, \downarrow} )\, \right] 
\;,
\label{eq:H_int}
\\
\mathcal{H}_{\mathrm{mix}} 
&=&
 \sum_{j=1}^M 
 \sum_{\sigma} 
 \left[\,  
 v_\mathrm{L}^{\phantom 0}  
           \,  a^{\dagger}_{\mathrm{L}, j,\sigma}  c^{\phantom{\dagger}}_{1,j, \sigma} 
\ + \ 
 v_\mathrm{R}^{\phantom 0}  
           \,  a^{\dagger}_{\mathrm{R}, j,\sigma}  c^{\phantom{\dagger}}_{N,j, \sigma} 
+ \mathrm{H.c.}             \,\right]   \;. 
\end{eqnarray} 
Here $c^{\dagger}_{i,j, \sigma}$ 
creates an electron with spin $\sigma$ at 
$\mbox{\boldmath $r$} = (i,j)$, and
 $n_{i,j, \sigma} = 
c^{\dagger}_{i,j, \sigma} c^{\phantom{\dagger}}_{i,j, \sigma}$.  
The periodic boundary condition is assumed along 
the $y$-direction  $c_{i,M+1, \sigma} \equiv c_{i,1, \sigma}$. 
In eq.\ (\ref{eq:E_d}),  $\epsilon_d$ is the bare onsite energy, 
and $\,\langle n_{i,j} \rangle_0 
\equiv \sum_{\sigma}\langle n_{i,j,\sigma} \rangle_0 $ 
is the non-magnetic solution of 
the Hartree-Fock approximation: 
$ \langle n_{i,j, \uparrow} \rangle_0 
=  \, \langle n_{i,j, \downarrow} \rangle_0$, 
this average generally depends on the position $(i,j)$.
The Hamiltonian for the two leads, 
$\mathcal{H}_{\rm L}$ and $\mathcal{H}_\mathrm{R}$, 
are written in terms of 
the one-particle energy $\epsilon_{\lambda k}^{\phantom{0}}$ and 
corresponding eigenfunction 
$\phi_{\lambda k}^{\phantom{0}}
(\mbox{\boldmath $r$}_{\lambda}^{\phantom{0}})$ 
for $\lambda=\mathrm{L},\,\mathrm{R}$. 
The operator $a_{\lambda, j, \sigma}^{\phantom{\dagger}} 
= \sum_k a_{\lambda k \sigma}^{\phantom{\dagger}} 
\phi_{\lambda k}^{\phantom{0}}
(\mbox{\boldmath $r$}_{\lambda,j}^{\phantom{0}})$ appearing 
in the mixing Hamiltonian $\mathcal{H}_\mathrm{mix}$ 
annihilates an electron 
at the interface $\mbox{\boldmath $r$}_{\lambda,j}^{\phantom{0}}$. 
We will be using units $\hbar=1$ unless otherwise noted.

The single-particle Green's function is defined by 
\begin{equation} 
G_{\mbox{\boldmath $r$}\mbox{\boldmath $r$}'}({\rm i}\varepsilon_n) 
 =  
-    \int_0^{\beta} \! {\rm d}\tau 
   \left \langle  T_{\tau} \,  
   c^{\phantom{\dagger}}_{\mbox{\boldmath $r$} \sigma} (\tau) 
   \, c^{\dagger}_{\mbox{\boldmath $r$}' \sigma} (0)     
                   \right \rangle  \, {\rm e}^{{\rm i}\, \varepsilon_n \tau} .
  \label{eq:G_Matsubara}                  
\end{equation} 
Here $\beta= 1/T$, $\varepsilon_n = (2n+1)\pi/\beta$, 
$c_{\mbox{\boldmath $r$} \sigma}^{\phantom{0}}(\tau) = 
{\rm e}^{\tau  {\cal H}} c_{\mbox{\boldmath $r$} \sigma}^{\phantom{0}} 
{\rm e}^{- \tau  {\cal H}}$  with $\mbox{\boldmath $r$}=(i,j)$, 
and $\langle \cdots \rangle$ denotes the thermal average 
$\mbox{Tr} \left[ \, {\rm e}^{-\beta  {\cal H} }\, {\cdots} 
\,\right]/\mbox{Tr} \, {\rm e}^{-\beta  {\cal H} }$. 
Since $U$ is acting only in the central region (C), 
the Dyson equation can be written as 
\begin{equation} 
  G_{\mbox{\boldmath $r$}\mbox{\boldmath $r'$}}({\rm i}\varepsilon_n)    
  =   G^0_{\mbox{\boldmath $r$}\mbox{\boldmath $r'$}}({\rm i}\varepsilon_n) 
    + \sum_{\mbox{\boldmath $r_1$},\mbox{\boldmath $r_2$} \in {\rm C}}
    \,G^0_{\mbox{\boldmath $r$}\mbox{\boldmath $r_1$}}({\rm i}\varepsilon_n)\, 
     \Sigma_{\mbox{\boldmath $r_1$}\mbox{\boldmath $r_2$}}^{\phantom{0}}
    ({\rm i}\varepsilon_n)  
   \, G_{\mbox{\boldmath $r_2$}\mbox{\boldmath $r'$}}
   ({\rm i}\varepsilon_n) \;.    
  \label{eq:Dyson}   
\end{equation}
Here $G^0_{\mbox{\boldmath $r$}\mbox{\boldmath $r'$}} 
({\rm i}\varepsilon_n)$ is 
the unperturbed Green's function 
corresponding to ${\cal H}^0$, 
which is determined with 
the non-magnetic solution of the Hartree-Fock 
approximation, so that 
\begin{equation}
\langle n_{\mbox{\boldmath $r$}}\rangle_0 
= 
- \frac{2}{\pi} \,  
\int^0_{-\infty} \! \mathrm{d} \omega\ 
\mathrm{Im}\, G^{0\,+}_{\mbox{\boldmath $r$}\mbox{\boldmath $r$}}
(\omega)\;. 
\label{eq:average} 
\end{equation}
The value of  $\langle n_{\mbox{\boldmath $r$}}\rangle_0$ 
is determined self-consistently 
from eqs.\ (\ref{eq:E_d}) and (\ref{eq:average}). 
To specify the various types of the Green's functions, 
we use the symbol $+$ ($-$) in the superscript 
as a label for the retarded (advanced) function, \textrm{i.e.}, 
$G_{\mbox{\boldmath $r$}\mbox{\boldmath $r'$}}^{\pm}(\omega) 
\equiv G_{\mbox{\boldmath $r$}\mbox{\boldmath $r'$}} 
(\omega \pm {\rm i}0^+)$. 
In eq.\ (\ref{eq:Dyson}), 
$\Sigma_{\mbox{\boldmath $r$}\mbox{\boldmath $r'$}}^{\phantom{0}}$ is 
the self-energy correction beyond the mean-field theory, and  
the summations with respect to $\mbox{\boldmath $r_1$}$ 
and $\mbox{\boldmath $r_2$}$  
run over the $N_{\rm C} = N \times M$ sites in the central region.
Therefore, the Dyson equation can be written 
in a matrix form, 
\begin{eqnarray}
\left\{ \widehat{\mbox{\boldmath ${\cal G}$}}(z) \right\}^{-1} &=&
\left\{ \widehat{\mbox{\boldmath ${\cal G}$}}^0(z) \right\}^{-1} 
 - \widehat{\mbox{\boldmath $\Sigma$}}(z)
\;,
  \label{eq:Dyson2}   
\\
\left\{ \widehat{\mbox{\boldmath ${\cal G}$}}^0(z) \right\}^{-1} 
&=&
z \, \widehat{\mbox{\boldmath $1$}}
 - \widehat{\mbox{\boldmath ${\cal H}$}}_{\rm C}^0 
- \widehat{\mbox{\boldmath ${\cal V}$}}_{\rm mix}(z) \;.
\label{eq:81}
\end{eqnarray}
Here
$\widehat{\mbox{\boldmath ${\cal G}$}}^0(z)$ 
is the  $N_{\rm C}\times N_{\rm C}$ matrix corresponding to 
the unperturbed Green's function   
$G^{0\,+}_{\mbox{\boldmath $r$}\mbox{\boldmath $r'$}}\,(\omega)\,$,
$\,\widehat{\mbox{\boldmath $1$}}$ is 
the $N_{\rm C} \times N_{\rm C}$ unit matrix,  
and 
\begin{eqnarray}
\widehat{\mbox{\boldmath $\mathcal{H}$}}_\mathrm{C}^0 &=&
\left[
\begin{matrix} 
 \mbox{\boldmath $h$}_1^0 & -t_x \mbox{\boldmath $1$} 
 & & \mbox{\boldmath $0$} \cr 
 -t_x \mbox{\boldmath $1$} & \mbox{\boldmath $h$}_2^0 & \ddots & \cr 
 & \ddots & \ddots & -t_x \mbox{\boldmath $1$} \cr 
 \mbox{\boldmath $0$} & & -t_x \mbox{\boldmath $1$} & 
 \mbox{\boldmath $h$}_N^0 \cr 
\end{matrix} 
\right]
\;,
\label{eq:85} 
\\
\widehat{\mbox{\boldmath ${\cal V}$}}_{\rm mix}(z) &=&
\left[
\begin{matrix}
v_{\rm L}^2 \mbox{\boldmath $F$}_{\rm L}(z) 
 & & & &  \cr
 & & & &  \cr
 & & \mbox{\boldmath $0$} & &  \cr
 & & & &  \cr
 & & & & 
 v_{\rm R}^2 \mbox{\boldmath $F$}_{\rm R}(z)
 \cr
\end{matrix}
\right],
\label{eq:87}
\\
\widehat{\mbox{\boldmath $\Sigma$}}(z) &=& 
\left[
\begin{matrix}
\mbox{\boldmath $\Sigma$}_{11}(z) & 
\mbox{\boldmath $\Sigma$}_{12}(z) & \ldots & 
\mbox{\boldmath $\Sigma$}_{1N}(z) \cr
\mbox{\boldmath $\Sigma$}_{21}(z) & \mbox{\boldmath $\Sigma$}_{22}(z) 
& \ddots & \vdots                  \cr
\vdots        & \ddots           & \ddots & \vdots  \cr
\mbox{\boldmath $\Sigma$}_{N1}(z) & \ldots 
  & \ldots & \mbox{\boldmath $\Sigma$}_{NN}(z) \cr
\end{matrix}
  \right].
\label{eq:86}
\end{eqnarray}
In eq.\ (\ref{eq:85}), 
$\mbox{\boldmath $1$}$ is the $M\times M$ unit matrix, and  
$\mbox{\boldmath $h$}_i^0$ is a tridiagonal $M\times M$ matrix 
consisting of $E_{i,j}$ at the $j$-th main diagonal part 
and $-t_y$ at sub- and superdiagonal parts. 
In eq.\ (\ref{eq:86}),
$\mbox{\boldmath $\Sigma$}_{ii'}(z)$ is 
a $M \times M$ matrix whose $(j,j')$ element 
is the self-energy $\Sigma_{\mbox{\boldmath $r$}\mbox{\boldmath $r'$}}$ 
for $\mbox{\boldmath $r$}=(i, j)$, $\mbox{\boldmath $r$}'=(i', j')$. 
In eq.\ (\ref{eq:87}), 
$\mbox{\boldmath $F$}_{\lambda}(z) = \left\{ F_{\lambda,jj'}(z) \right\}$  
is the Green's function 
at the interface of the isolated lead at $\lambda = \mathrm{L}, \mathrm{R}$; 
\begin{equation}
F_{\lambda,jj'}(z) \, = \, \sum_k 
\frac{
\phi_{\lambda k}^{\phantom{0}}
(\mbox{\boldmath $r$}_{\lambda,j}^{\phantom{0}}) 
\phi_{\lambda k}^{*}
(\mbox{\boldmath $r$}_{\lambda,j'}^{\phantom{0}})} 
{z - \epsilon_{\lambda k}^{\phantom{0}} } \;.
\end{equation}
From this and the hybridization $v_{\lambda}^{\phantom{0}}$,
the level width due to the mixing with the leads 
is given by $\mbox{\boldmath $\Gamma$}_{\lambda}(\omega)
\equiv -v_{\lambda}^2\, 
\mbox{Im}\, \mbox{\boldmath $F$}_{\lambda}^+(\omega)$
for $\lambda = \mathrm{L}$ and $\mathrm{R}$. 
In the present study, we assume a diagonal form 
$\mbox{\boldmath $F$}_{\rm L}^+(\omega)
=\mbox{\boldmath $F$}_{\rm R}^+(\omega)
= -{\rm i}\,\pi \rho \,\mbox{\boldmath $1$}$
taking the local density of states 
at the interfaces $\rho$ to be a constant 
independent of $\omega$, so that 
the effects of the mixing can be parameterized by 
$\Gamma_{\lambda} = \pi \rho \, v^2_{\lambda}$. 
Note that, 
because of the periodic boundary condition 
in the $y$-direction, 
the $M \times M$  matrix functions 
can be diagonalized as 
$
\mbox{\boldmath $\Sigma$}_{ii'} 
 = \sum_{m=1}^M 
\mbox{\boldmath $\chi$}_m  
\Sigma_{ii'}^{(m)}
\mbox{\boldmath $\chi$}_m^{\dagger}
$.
Here $\mbox{\boldmath $\chi$}_m$ is the eigenvector for the subbands,  
and the corresponding eigenvalue is given by 
  $\epsilon_m = -2 t_y \cos (2\pi m/M)$ for $m=1,2,\ldots, M$.

At zero temperature, the dc conductance can be calculated 
from the Green's function as
\begin{equation}
 g =  \frac{2 e^2}{h} \, 
               \mbox{Tr} \left[\,4\,
               \mbox{\boldmath $\Gamma$}_{\rm R} (0) \, 
               \mbox{\boldmath $G$}_{N1}^{+}(0)\, 
               \mbox{\boldmath $\Gamma$}_{\rm L}(0) \,  
               \mbox{\boldmath $G$}_{1N}^{-}(0)  
               \,\right] , 
\label{eq:g_multi}
\end{equation}
where $\mbox{\boldmath $G$}_{ii'}(z)$ is a $M\times M$ matrix 
that corresponds to the $(i,i')$ partitioned part of 
$\widehat{\mbox{\boldmath ${\cal G}$}}(z)$. 
Furthermore, at $T=0$,
the total charge displacement $\Delta N_{\rm tot}$ 
can be calculated from the Friedel sum rule,\cite{LangerAmbegaokar,YT} 
as 
\begin{eqnarray}
& &
\Delta N_{\rm tot} 
= 
\frac{1}{\pi {\rm i}}\, 
\log [\, \det 
 \mbox{\boldmath $S$}
\,] \;,
\label{eq:Friedel}
\\
\nonumber
\\
& &\mbox{\boldmath $S$}
\, = \,  
  \left [\, 
 \begin{matrix} 
          \mbox{\boldmath $1$}  & \mbox{\boldmath $0$}  \cr
           \mbox{\boldmath $0$}  & \mbox{\boldmath $1$}  \cr  
 \end{matrix}
  \,\right ] 
- \, 2 {\rm i} 
  \left [ 
 \begin{matrix} 
          \mbox{\boldmath $\Gamma$}_{\rm L}(0)  & \mbox{\boldmath $0$}  \cr
           \mbox{\boldmath $0$}  & \mbox{\boldmath $\Gamma$}_{\rm R}(0)  \cr  
 \end{matrix}
  \right ]  
  \left [ 
 \begin{matrix}
             \mbox{\boldmath $G$}_{11}^{+}(0)  
           & \mbox{\boldmath $G$}_{1N}^{+}(0)  \cr 
             \mbox{\boldmath $G$}_{N1}^{+}(0)  
           & \mbox{\boldmath $G$}_{NN}^{+}(0)  \cr  
 \end{matrix}
  \right ]  
     .   
\nonumber     
\\ 
\label{eq:S}
\end{eqnarray}
Specifically, in the case of the constant density of states $\rho$,   
the Anderson compensation theorem holds,\cite{Anderson} 
and then the charge displacement of whole the system 
coincides with the number of electrons in the central region:~\cite{YT} 
\begin{equation} 
\Delta N_{\rm tot} \, = \, 
 \sum_{i=1}^N \sum_{j=1}^M \, \langle n_{i,j}\rangle \;. 
\end{equation}

These expressions, eqs.\ (\ref{eq:g_multi}) and (\ref{eq:Friedel}),
show that at $T=0$ the conductance and charge displacement can be  
determined by the value of the Green's function 
at the Fermi energy $\omega=0$. 
Therefore, one can introduce a noninteracting Hamiltonian 
that leads exactly the same $g$ and $\Delta N_{\rm tot}$ of 
the interacting system.\cite{ao7-ao9}
This is due to the property 
$\mbox{Im}\,\widehat{\mbox{\boldmath $\Sigma$}}^+(0) 
 = \widehat{\mbox{\boldmath $0$}}$ at $T=0$,
and in the central region such an effective Hamiltonian 
can be constructed as
\begin{equation} 
 \widehat{\mbox{\boldmath ${\cal H}$}}_{\rm C}^{\rm eff} 
\, = \, \widehat{\mbox{\boldmath ${\cal H}$}}_{\rm C}^0 
+ \mbox{Re}\, \widehat{\mbox{\boldmath $\Sigma$}}^+(0) 
\;.
\label{eq:effective}
\end{equation} 
Note that the self-energy 
is defined with respect to the whole system, 
so that $\mbox{Re}\,\widehat{\mbox{\boldmath $\Sigma$}}^+(0)$ depends 
not only on $U$ but also on
$\Gamma_{\rm L}^{\phantom{0}}$ and $\Gamma_{\rm R}^{\phantom{0}}$ 
through the unperturbed Green's function 
$\widehat{\mbox{\boldmath ${\cal G}$}}^0(z)$.
We see in the next section that
the eigenvalues of 
$\widehat{\mbox{\boldmath ${\cal H}$}}_{\rm C}^{\rm eff}$ 
correspond approximately to the peaks of the resonant states.

\section{the Second Order Perturbation Theory}

\begin{figure}[tb]
\begin{center}
\leavevmode
\includegraphics[width=0.5\linewidth, clip, 
trim = 4cm 9.5cm 4cm 11cm]{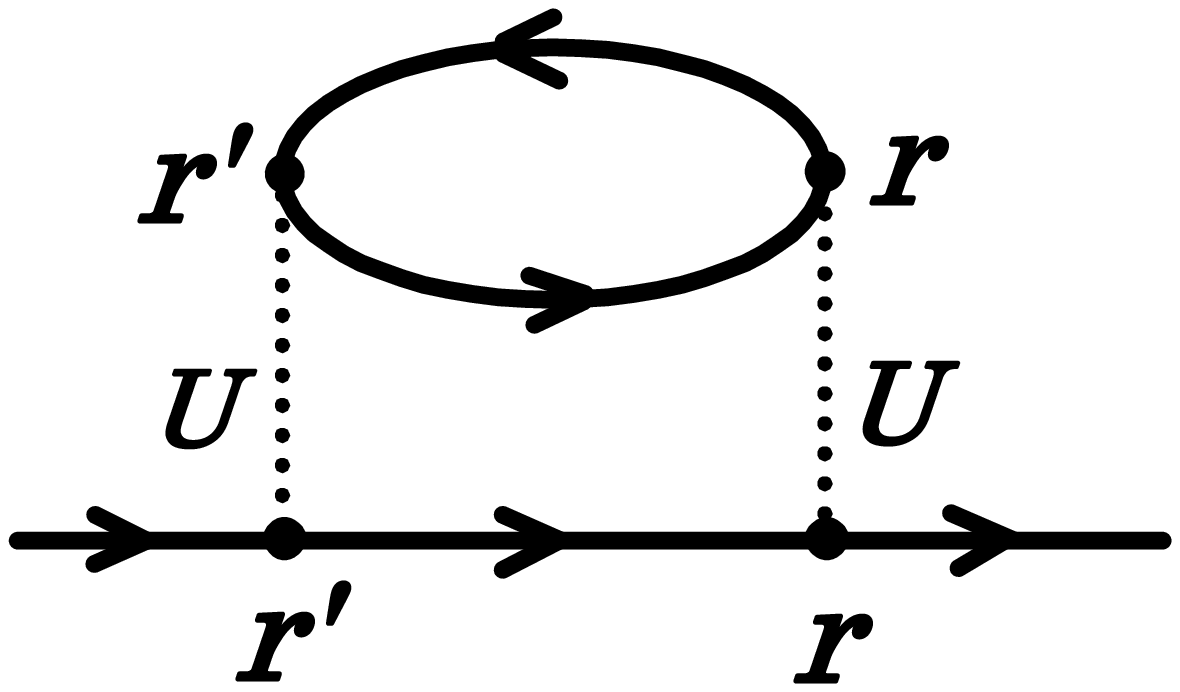}
\caption{The order $U^2$ self-energy 
$\Sigma_{\mbox{\boldmath $r$}\mbox{\boldmath $r'$}}
(\mathrm{i}\varepsilon$).} 
\label{fig:fig02}
\end{center}
\end{figure}

In this section, we calculate  
the self-energy  $\mbox{Re}\,\widehat{\mbox{\boldmath $\Sigma$}}^+(0)$ up to 
the second order with respect to $\mathcal{H}_\mathrm{C}^\mathrm{int}$ 
defined by eq.\ (\ref{eq:H_int}), 
and show the results for the conductance and charge displacement. 
In order to determine the unperturbed Green's 
function $G^{0}_{\mbox{\boldmath $r$}\mbox{\boldmath $r'$}}$  
with the Hartree-Fock approximation 
assuming the paramagnetic ground state,
we calculate first the value of $E_{i,j}$ by 
solving eqs.\ (\ref{eq:E_d}) and (\ref{eq:average}) 
for given $\epsilon_d$ and $U$. 
Then, we take into account the lowest order self-energy correction due 
to the fluctuation $\mathcal{H}_\mathrm{C}^\mathrm{int}$.  
It is described by the diagram shown in Fig.\ \ref{fig:fig02}, 
and its value at $T=0$ and $\omega=0$ is given by  
\begin{eqnarray}
 & &
 \Sigma_{\mbox{\boldmath $r$}\mbox{\boldmath $r'$}}^+(0)  \,  = 
   \,  - \, U^2   \int_{-\infty}^{\infty} \! \int_{-\infty}^{\infty} 
         \frac{{\rm d}\varepsilon\, {\rm d}\varepsilon'}{(2\pi)^2}
       \, 
       G^{0}_{\mbox{\boldmath $r$}\mbox{\boldmath $r'$}}
       ({\rm i} \varepsilon)  \,  
          G^{0}_{\mbox{\boldmath $r$}\mbox{\boldmath $r'$}}
          ({\rm i} \varepsilon') \, 
          G^{0}_{\mbox{\boldmath $r'$}\mbox{\boldmath $r$}}
          ({\rm i} \varepsilon + {\rm i} \varepsilon') 
\;.
\label{eq:Self_2v} 
\end{eqnarray}
The explicit form of $G^{0}_{\mbox{\boldmath $r$}\mbox{\boldmath $r'$}}$  
can be derived from eq.\ (\ref{eq:81}).
Note that the retarded function at $\omega=0$ and $T=0$ 
can be obtained from the Matsubara function, \textrm{i.e.}, 
$\Sigma_{\mbox{\boldmath $r$}\mbox{\boldmath $r'$}}^+(0) 
= \Sigma_{\mbox{\boldmath $r$}\mbox{\boldmath $r'$}} 
({\rm i}\varepsilon)|_{\varepsilon \to 0^+}$. 
From eq.\ (\ref{eq:Self_2v}), 
one can also confirm explicitly the property 
$\mbox{Im}\, \Sigma_{\mbox{\boldmath $r$}\mbox{\boldmath $r'$}}^+(0)=0$, 
at $\omega=0$ and $T=0$. 
We calculate all the $N_{\rm C} \times N_{\rm C}$ elements 
of $\mbox{Re}\, \Sigma_{\mbox{\boldmath $r$}\mbox{\boldmath $r'$}}^+(0)$ 
carrying out the integrations numerically. 
Then, we obtain the full Green's function 
$G_{\mbox{\boldmath $r$}\mbox{\boldmath $r'$}}^{+}(0)$ 
substituting
 $\mbox{Re}\, \Sigma_{\mbox{\boldmath $r$}\mbox{\boldmath $r'$}}^+(0)$ 
into the Dyson equation eq.\ (\ref{eq:Dyson2}).

In the previous work we have used this method to 
study the ground state properties at half-filling,\cite{YT} 
where the spatial distribution of the charge is uniform 
independent of the details of the parameters.
In this case the perturbation theory works rather well.~\cite{ao7-ao9}
Specifically, for the single Anderson impurity model, 
which corresponds to a special limit $N_\mathrm{C}=1$, 
the order $U^2$ self-energy describes the low-energy 
Fermi-liquid behavior properly\cite{YamadaYosida} 
and gives the spectral function that is 
qualitatively correct even for rather large $U$.\cite{Hewson}
Away from half-filling, however, 
the spatial distribution of the charge is inhomogeneous, 
and it makes the applicability of the second-order perturbation 
theory rather worse.  
As the density functional theory\cite{HohenbergKohn}
teaches us, in order to get the ground state correctly, 
one needs the method that can reproduce the charge distribution accurately.
We know the correct charge distribution at half-filling 
but we do not away from half-filling. 
Therefore, away from half-filling the charge distribution is the quantity 
that should be calculated.
This makes the situation away from half-filling different from 
that at half-filling. 
For the Anderson impurity, 
the second-order perturbation theory 
has been applied to away from half-filling,\cite{horva1} 
and a modified method has also been examined 
by several authors.~\cite{Yeyati,Saso} 
Horvati\'c and Zlati\'c has shown for the Anderson model that
the radius of the convergence of the 
perturbation expansion with respect to $U$ decreases 
rapidly when the system goes away from half-filling.~\cite{horva1} 
We see that there is a similar tendency also for $N_\mathrm{C}>1$ in 
the following subsection. 
To check the applicability of the approximation, 
we calculate the total charge in the interacting region 
with two different methods. 
The first one is using the integration of the local density of states (DOS)
up to the Fermi level, 
\begin{equation}
 \sum_{i=1}^N \sum_{j=1}^M 
 \, \langle n_{i,j}\rangle 
 \, = \, 
- \frac{2}{\pi} \,  
\int^0_{-\infty} \mathrm{d} \omega\,
\, 
\mathrm{Tr}\ \mathrm{Im}  
\left[ \widehat{\mbox{\boldmath ${\cal G}$}}^+(\omega)
\right]
\; 
\label{eq:average2} 
\end{equation}
where $\widehat{\mbox{\boldmath ${\cal G}$}}^+(\omega)$ 
should be calculated using the order $U^2$ self energy.
The second method is using the Friedel sum rule 
eqs.\ (\ref{eq:Friedel}) and (\ref{eq:S}), and thus 
the charge is determined by the value at the Fermi energy $\omega=0$. 
The two results agree well for small $U$, 
and it gives some insights into the applicability.

In the following subsections, 
we apply the formulation to the Anderson impurity in \S \ref{subsec:0d} 
in order to demonstrate how it works in the simplest case.  
Then, we examine the one-dimensional chain in \S \ref{subsec:1d} 
and the two-dimensional lattice in \S \ref{subsec:2d}. 
Specifically, we concentrate on 
the case $\Gamma_{\rm L} = \Gamma_{\rm R}$ ($\equiv\ \Gamma$),
where the system has the inversion symmetry.

\subsection{Anderson impurity}
\label{subsec:0d}

We consider here the single Anderson impurity 
that corresponds to the case of $N_\mathrm{C} =1$ ($M=1$, $N=1$). 
In this subsection,  
we use the width of the resonance peak for the noninteracting case 
as the unit of the energy, i.e., $\Delta = 2\Gamma$.

\begin{figure}[tb]
\begin{center}
\leavevmode
\includegraphics[width=0.6\linewidth, clip, 
trim = 2cm 14cm 3cm 2.5cm]{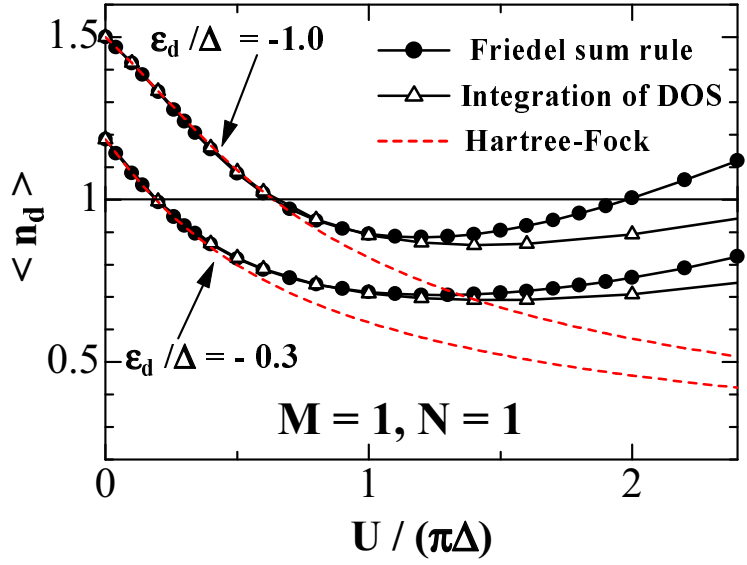}
\caption{
The number of electrons at the impurity site $\langle n_d \rangle$
of the Anderson model ($M=1$, $N=1$),  
where $\epsilon_d / \Delta$ is taken to be $-0.3$ and $-1.0$.
The results are obtained with the two different methods which use 
($\bullet$) the Friedel sum rule and 
($\vartriangle$) the integration of DOS up to the Fermi level. 
The two dashed lines are the results obtained with 
the Hartree-Fock approximation  
for the same parameters: $\epsilon_d / \Delta = -0.3$ and $-1.0$. 
} 
\label{fig:fig11}
\end{center}
\end{figure}

In Fig.\ \ref{fig:fig11}, 
the average number of electrons 
at the impurity site $\langle n_d\rangle$ is 
shown as a function of $U$  for $\epsilon_d/\Delta = -0.3$ and $-1.0$. 
The system has the electron-hole symmetry 
at the value of $U$ where $\langle n_d\rangle$ is equal to 1.0. 
For $U/(\pi\Delta) \lesssim 1.0$, 
the results obtained from the Friedel sum rule ($\bullet$) 
agree well with that of the integration of DOS ($\triangle$).  
The range of $U$ where we have good agreement depends on $\epsilon_d$,
but the dependence is weak for small $\epsilon_d$. 
When the onsite energy $\epsilon_d$ is unchanged, 
the local charge $\langle n_d\rangle$ must be a decreasing function 
of $U$ and converges to a finite constant in the $U \to \infty$.
Therefore, the upturn seen in Fig.\ \ref{fig:fig11} 
indicates the breakdown of the approximation for large $U$.
In Fig.\ \ref{fig:fig11}, we have also shown   
the Hartree-Fock results (dashed lines) 
for $\epsilon_d / \Delta = -0.3$ and $-1.0$.
For  $U< -2\epsilon_d$ (or equivalently $\langle n_d \rangle > 1.0$),  
the Hartree-Fock result is larger slightly than that 
of the second-order perturbation result, 
although it might be hard to see this feature in the figure.
In contrast, for $U> -2\epsilon_d$ 
(or equivalently $\langle n_d \rangle < 1.0$),
the Hartree-Fock results is smaller than that of the perturbation 
results, and the difference increases with $U$.
These results are consistent with that in 
the seminal work by Horvati\'c and Zlati\'c,~\cite{horva1} 
and suggest that the method is applicable 
for $U/(\pi\Delta) \lesssim 1.0$.

\begin{figure}[tb]
\begin{center}
\leavevmode
\includegraphics[width=0.6\linewidth, clip, 
trim = 1.5cm 0cm 2cm 0cm]{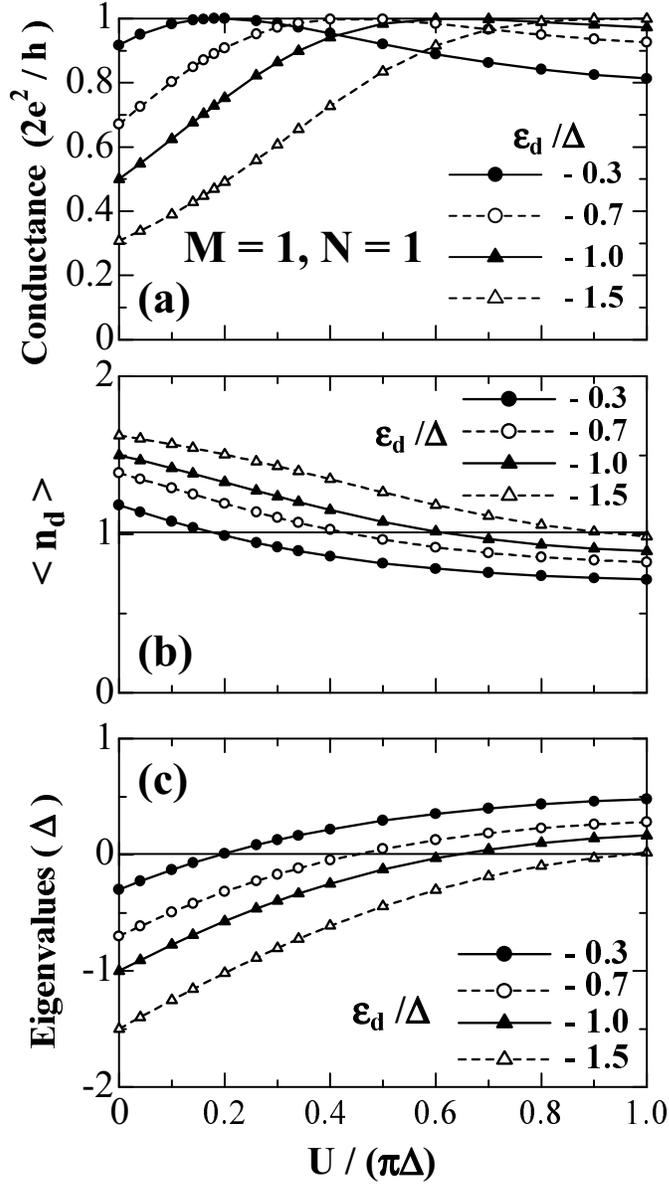}
\caption{
Ground state properties of the single Anderson model:  
(a) conductance,  
(b) local charge $\langle n_d \rangle$, and 
(c) $\widetilde{E}_d 
\equiv \epsilon_d  + U\,\langle n_d\rangle_0/2 + \mathrm{Re}\, \Sigma^+(0)$
which corresponds to the eigenvalue 
of $\widehat{\mbox{\boldmath ${\cal H}$}}_{\rm C}^{\rm eff}$. 
Here the onsite energy $\epsilon_d / \Delta$ is taken 
to be ($\bullet$) $-0.3$, ($\circ$) $-0.7$, ($\blacktriangle$) $-1.0$,  
and ($\triangle$) $-1.5$.
}
\label{fig:fig21}
\end{center}
\end{figure}

In Fig.\ \ref{fig:fig21}, the result of the conductance 
is shown  as a function of $U/(\pi\Delta)$ for several values 
of $\epsilon_d/\Delta = -0.3,\, -0.7,\, -1.0$, and $-1.5$.
The conductance shows a maximum at half-filling 
$\langle n_d\rangle=1$, which corresponds 
to $U=-2\epsilon_d$.
This is because at half-filling
the resonance level is situated exactly at the Fermi level. 
The change of the position of the resonance state can be traced 
through the eigenvalue of the effective 
Hamiltonian $\widehat{\mbox{\boldmath ${\cal H}$}}_{\rm C}^{\rm eff}$. 
For the Anderson model the eigenvalue is given by 
$\widetilde{E}_d \equiv 
\epsilon_d + U\,\langle n_d\rangle_0/2 + \mathrm{Re}\, \Sigma^+(0)$, 
which is plotted as a function of $U$ in Fig.\ \ref{fig:fig21} (c). 
The perfect transmission occurs, 
when the eigenvalue crosses the Fermi level at $\omega=0$. 
In the case of the Anderson impurity,
$\widetilde{E}_d$ corresponds exactly to the position of the resonance state. 
However, for larger systems consisting of a number of 
the interacting sites $N_{\rm C}>1$, 
the precise position of the resonant peaks deviates 
slightly from the eigenvalue  
of $\widehat{\mbox{\boldmath ${\cal H}$}}_{\rm C}^{\rm eff}$  
because of the additional effect of the hybridization  
$\widehat{\mbox{\boldmath ${\cal V}$}}_{\rm mix}$ 
appearing in the right-hand side of eq.\ (\ref{eq:81}).

\subsection{One-dimensional chain}
\label{subsec:1d}

We next consider one-dimensional chains 
connected to noninteracting leads, 
which is the case of  $M \equiv 1$ and $N \geq 2$. 
The system can be regarded as a model for a series of quantum dots 
or the atomic chain of nanometer size. 
In the electron-hole symmetric case
the second-order perturbation theory described above 
has already been applied to this model.~\cite{ao7-ao9} 
The results obtained at half-filling show 
the even-odd oscillatory behavior:  
the perfect transmission due to 
resonant tunneling occurs for odd $N$, 
while for even $N$ the conductance decreases with increasing $N$ 
showing a tendency towards the development of a Mott-Hubbard insulator gap.
In this subsection, 
we study the ground state properties away from half-filling. 
In the following, 
we use the hopping matrix element $t$, instead of $\Delta$, 
as the energy scale, 
and take the mixing parameter 
to be $\Gamma/t = 0.75$ in the numerical calculations.

To check the applicability of the perturbation theory 
in the one-dimensional case, 
we have compared the number of electrons 
in the interacting region $\Delta N_{\rm tot}$ with 
the Friedel sum rule and the integration of DOS up to the Fermi level
as we have done for the Anderson impurity in the above. 
Specifically, we have examined  
the chain of length $N=3$ and $N=4$ 
taking the onsite energy to be $\epsilon_d/t = -0.3$ and $-1.0$. 
Although the results are not shown here, 
we have obtained the figures, 
the feature of which is quite similar to that of Fig.\ \ref{fig:fig11}. 
The results of the two different methods agree well 
at $U/(2 \pi t) \lesssim 0.7\,$  both for $N=3$ and $N=4$. 
Therefore, in this subsection we discuss the results obtained 
in the small $U$ region;  $U/(2 \pi t) \lesssim 0.7$.

\begin{figure}[tb]
\begin{center}
\leavevmode
\includegraphics[width=0.6\linewidth, clip, 
trim = 1.5cm 0cm 2cm 0cm]{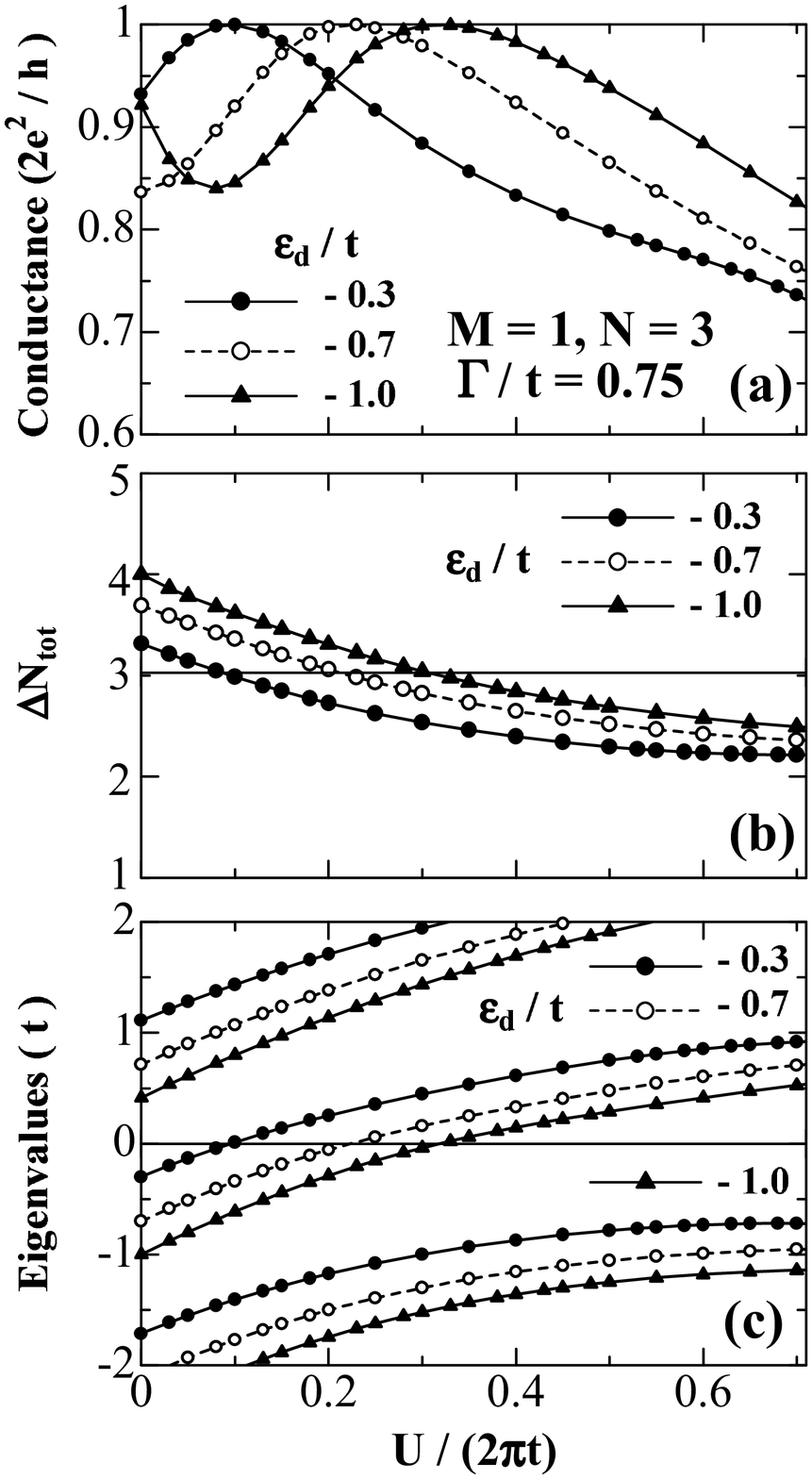}
\caption{
Ground state properties of a one-dimensional Hubbard model 
of odd number sites ($M=1$, $N=3$): 
(a) conductance, 
(b) 
local charge in the interacting region 
$\sum_{i=1}^N \sum_{j=1}^M \, \langle n_{i,j}\rangle$, 
and 
(c) eigenvalues of  
$\widehat{\mbox{\boldmath ${\cal H}$}}_{\rm C}^{\rm eff}$.  
Here $\epsilon_d / t$ is taken to be  ($\bullet$) $-0.3$, 
 ($\circ$) $-0.7$, and  ($\blacktriangle$) $-1.0$. 
}
\label{fig:fig22}
\end{center}
\end{figure}

\begin{figure}[tb]
\begin{center}
\leavevmode
\includegraphics[width=0.6\linewidth, clip, 
trim = 1.5cm 0cm 2cm 0cm]{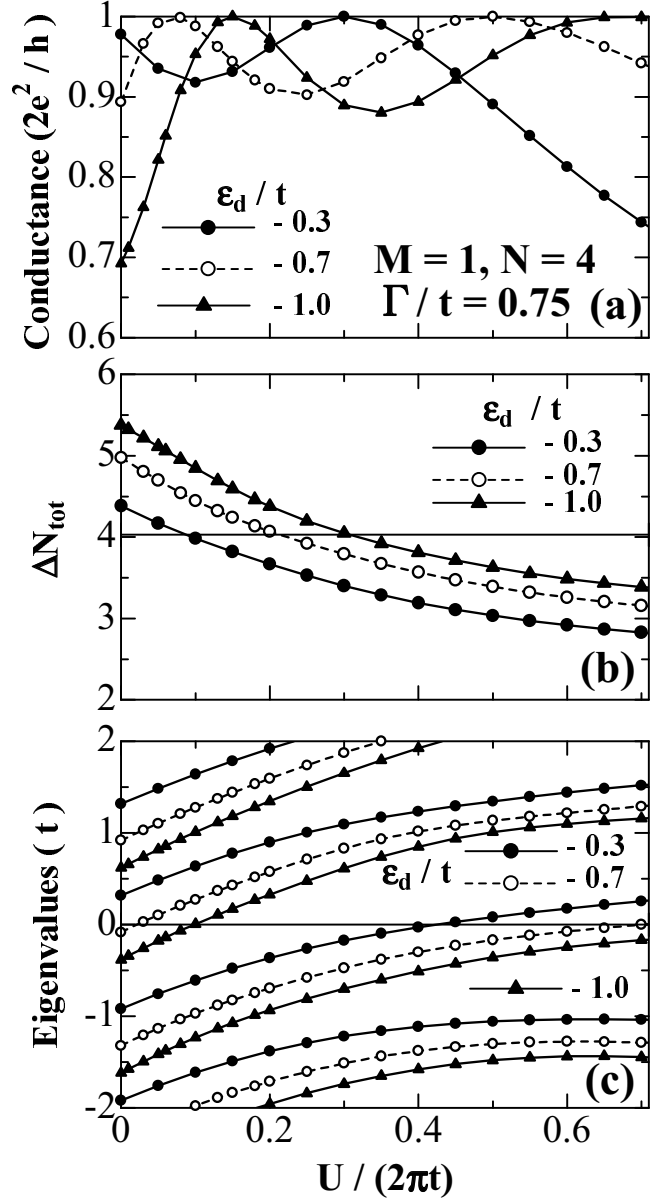}
\caption{
Ground state properties of a one-dimensional Hubbard model 
of even number sites ($M=1$, $N=4$): 
(a) conductance, 
(b) 
local charge in the interacting region
 $\sum_{i=1}^N \sum_{j=1}^M \, \langle n_{i,j}\rangle$, 
and 
(c) eigenvalues of  $\widehat{\mbox{\boldmath ${\cal H}$}}_{\rm C}^{\rm eff}$. 
Here $\epsilon_d / t$ is taken to be  ($\bullet$) $-0.3$, 
 ($\circ$) $-0.7$, and  ($\blacktriangle$) $-1.0$. 
}
\label{fig:fig23}
\end{center}
\end{figure}

\begin{figure}[tb]
\begin{center}
\leavevmode
\includegraphics[width=0.6\linewidth, clip, 
trim = 1.0cm 14cm 1.5cm 0.5cm]{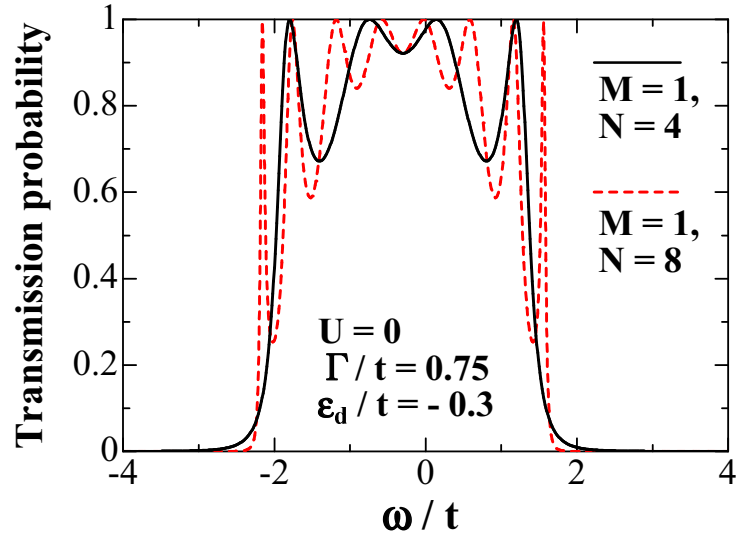}
\caption{Transmission probability 
vs $\omega$ for noninteracting chains ($M=1$),
where $\epsilon_d/t=-0.3$ and $\Gamma/t=0.75$.
The solid line is for the system of length $N=4$,
 and the dashed line is for $N=8$.
 }
\label{fig:fig19}
\end{center}
\end{figure}

In Figs.\ \ref{fig:fig22} and \ref{fig:fig23}, 
the ground state properties  
of the chains of length $N=3$ and $N=4$ 
are shown, respectively, where 
the onsite energy $\epsilon_d/t$ is taken to be 
($\bullet$) $-0.3$, ($\circ$) $-0.7$, 
and  ($\blacktriangle$) $-1.0$. 
The system of $N=3$ and $4$ can be regarded as 
the simplest examples of the odd and even chains, respectively, 
which capture the essence of the one-dimensional network. 
Note that the system has the electron-hole symmetry 
when $U + 2\,\epsilon_d =0$.
It corresponds to the value of $U$, 
at which the local charge becomes $\Delta N_{\rm tot}=N$ 
and the conductance shows 
a maximum (minimum) for odd (even) $N$. 
When an eigenvalue in the middle, which is shown 
in the panel (c) of Figs.\ \ref{fig:fig22} and \ref{fig:fig23}, 
crosses the Fermi level, the conductance shows a peak. 
For $N=3$, this happens only at half-filling 
in the situations shown in Fig.\ \ref{fig:fig22} (c), 
and the position of the conductance peak coincides exactly 
with the intersection of the central eigenvalue and the Fermi level.
In the case of $N=4$, the conductance 
shows some peaks away from half-filling.
We see two peaks in the curves 
for ($\circ$) $\epsilon_d /t=-0.7$ and ($\blacktriangle$) $-1.0$ 
in Fig.\ \ref{fig:fig23} (a).
These peaks in each of the curves 
correspond to the eigenvalues which cross the Fermi 
level at $\omega=0$ successively in Fig.\ \ref{fig:fig23} (c).
As mentioned in the above, 
the position of the resonance peaks  
deviates from that of the eigenvalue 
because of the energy-shift due to the hybridization. 
The conductance for  ($\bullet$) $\epsilon_d /t=-0.3$ 
in Fig.\ \ref{fig:fig23} (c) 
shows a single peak
because only one eigenvalue among the four crosses the Fermi level. 
In the noninteracting case 
this feature can be seen also in the $\omega$ dependence of the 
transmission probability, which is  
shown in Fig.\ \ref{fig:fig19} for $\epsilon_d /t = -0.3$.
In the solid line (for $N=4$) two peaks among the four are situated below 
the Fermi level $\omega=0$, and they move towards 
the Fermi level when $U$ is switched on.~\cite{ao10}
These examples show that 
the eigenvalues contain information 
about resonant states  situated near the Fermi level. 
We note that away from half-filling 
not only the off-diagonal part\cite{ao7-ao9}
but the diagonal part of 
 $\Sigma_{\mbox{\boldmath $r$}\mbox{\boldmath $r'$}}^+(0)$ is finite 
 and plays an important role on the charge distribution. 
Furthermore, we note that an analogous description has been used 
in the translational invariant systems, for which the self-energy 
can be diagonalized in the  $\mbox{\boldmath $k$}$ (wave-vector) space 
and the dispersion for the quasi-particles has a correction 
through the $\mbox{\boldmath $k$}$ dependence of the self-energy.

\begin{figure}[tb]
\begin{center}
\leavevmode
\includegraphics[width=0.6\linewidth, clip, 
trim = 1.0cm 3cm 1.5cm 1cm]{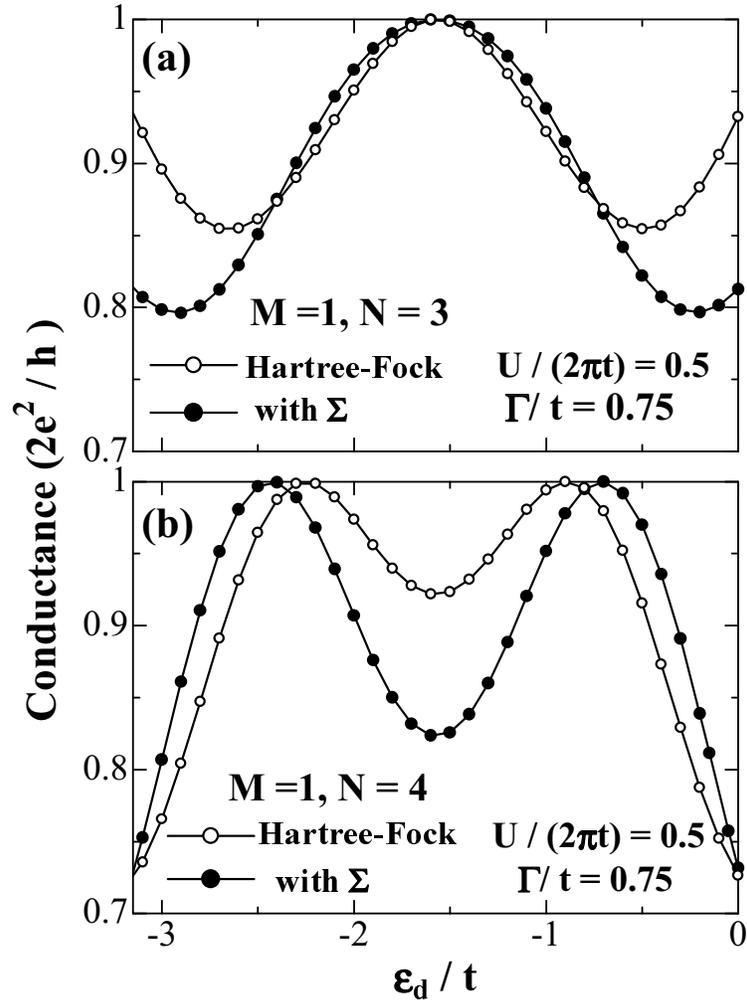}
\caption{The $\epsilon_d$ dependence of the conductance 
through the one-dimensional chains ($M=1$) of length 
(a) $N=3$, and (b) $N=4$.
The solid circles ($\bullet$)
 denote the results obtained using the second-order 
self-energy corrections, and the open circles ($\circ$) 
denotes the Hartree-Fock results.
The Coulomb interaction is taken to be $U/(2 \pi t) = 0.5$,
so that the electron-hole symmetry holds 
at $\epsilon_d = -U/2$ ($\simeq -1.6\, t$). 
}
\label{fig:fig24}
\end{center}
\end{figure}

So far, we have discussed mainly the $U$ dependence of the conductance and 
other quantities. In real quantum dot systems, however, 
the onsite energy  $\epsilon_d$ can be controlled by the gate voltage. 
Therefore, the $\epsilon_d$ dependence should also be examined.
In Fig.\ \ref{fig:fig24}  
the conductance for (a) $N=3$ and (b) $N=4$ is shown  
as a function of $\epsilon_d$ taking the Coulomb interaction 
to be $U/(2 \pi t)=0.5$. 
The conductance is symmetric with respect to  
the half-filled point $\epsilon_d = -U/2$ ($\simeq -1.6\, t$)
because of the electron-hole symmetry of the model.
To see the effects of the self-energy correction due to 
the residual interaction, in the figures we have also shown 
the Hartree-Fock results ($\circ$) obtained using the
unperturbed Green's function. 
In the case of $N=3$, the conductance shows a peak at 
half-filling, and the self-energy correction makes it broader.  
This is due to the Kondo effect, 
and is consistent with the numerical 
renormalization group results~\cite{Izumida4-5} 
for the Anderson impurity.
On the other hand, in the case of $N=4$, 
the self-energy correction makes 
the valley of the conductance around the half-filling deep and wide. 
This can be explained as a tendency towards 
the development of a Mott-Hubbard insulator gap.

\subsection{Two-dimensional lattice}
\label{subsec:2d}

In this subsection we consider the two-dimensional Hubbard model, 
as a model for a quantum dot superlattice. 
In the numerical calculations, 
we take the hopping matrix elements to be isotropic $t_x=t_y$ ($\equiv t$), 
and take the mixing parameter to be $\Gamma/t = 0.75$. 
Specifically, we examine a system of the size $M=4$, $N=8$: 
the number of the interacting sites is $N_{\rm C}=4 \times 8$.

\begin{figure}[tb]
\begin{center}
\leavevmode
\includegraphics[width=0.6\linewidth, clip, 
trim = 2.5cm 3cm 0cm 1cm]{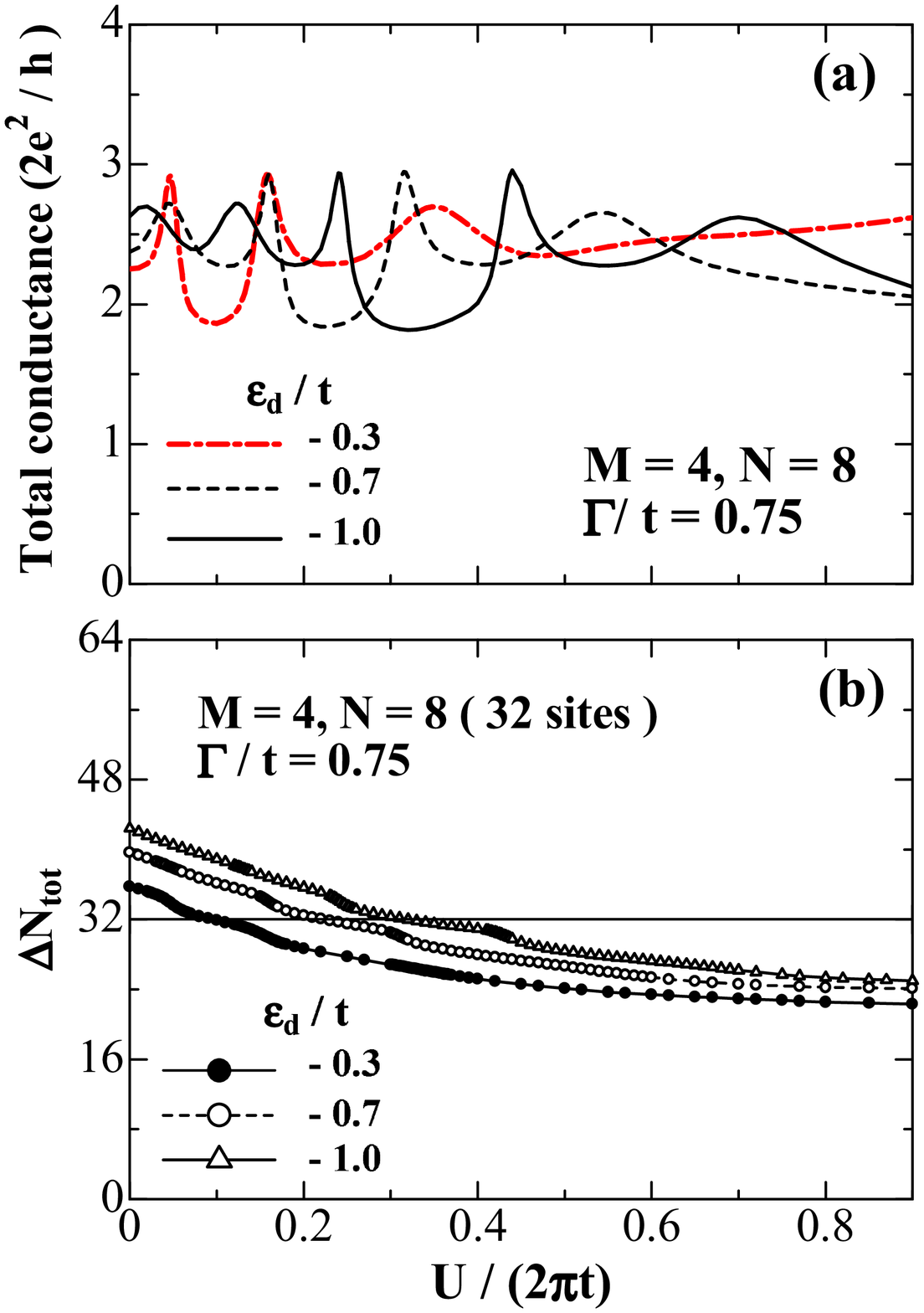}
\caption{
Ground state properties of 
the two-dimensional Hubbard model of the size $M=4$, $N=8$, 
(a) total conductance, and (b) 
local charge in the interacting region
 $\sum_{i=1}^N \sum_{j=1}^M \, \langle n_{i,j}\rangle$. 
Here $\epsilon_d / t$ is taken to be  ($\bullet$) $-0.3$, 
 ($\circ$) $-0.7$, and  ($\blacktriangle$) $-1.0$. 
}
\label{fig:fig41}
\end{center}
\end{figure}

\begin{figure}[tb]
\begin{center}
\leavevmode
\includegraphics[width=0.7\linewidth, clip, 
trim = 1.5cm 0cm 1cm 0cm]{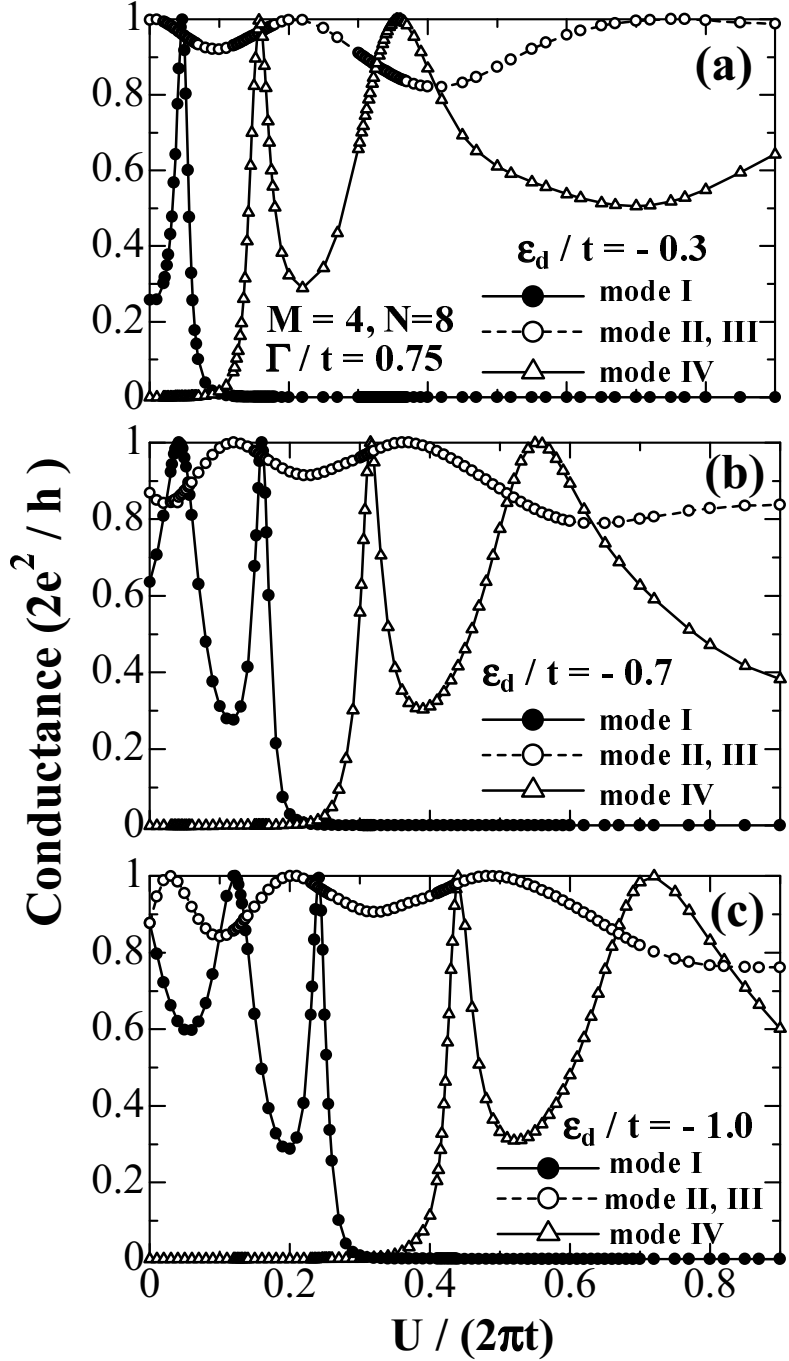}
\caption{Contributions of each subband on the 
conductance of the two-dimensional cluster of the size ($M=4$, $N=8$). 
Here, $\epsilon_d / t$ is taken to be 
 (a) $-0.3$, (b) $-0.7$, and (c) $-1.0$. }
\label{fig:fig40}
\end{center}
\end{figure}

\begin{figure}[tb]
\begin{center}
\leavevmode
\includegraphics[width=0.6\linewidth, clip, 
trim = 1.5cm 14cm 1cm 1cm]{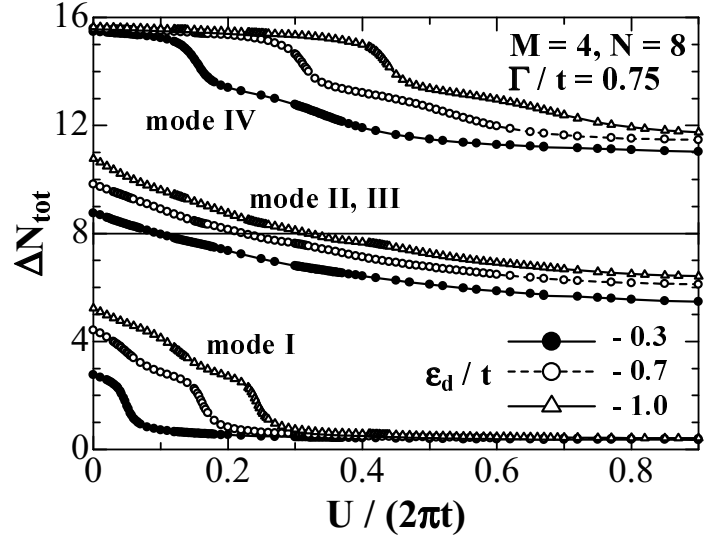}
\caption{
Contributions of each subband on the 
local charge in the interacting region
 $\sum_{i=1}^N \sum_{j=1}^M \, \langle n_{i,j}\rangle$, 
of the two-dimensional cluster of the size ($M=4$, $N=8$). 
Here, $\epsilon_d / t$ is taken to be ($\bullet$) $-0.3$,
($\circ$) $-0.7$, and ($\triangle$) $-1.0$. 
}
\label{fig:fig30}
\end{center}
\end{figure}

In Fig.\ \ref{fig:fig41}, the conductance and the number of electrons 
in the interacting region are shown 
for several values of $\epsilon_d/t =-0.3$, $-0.7$, and $-1.0$. 
In the lower panel (b) we see that the local charge 
$\Delta N_{\rm tot}$,  which is obtained using the Friedel sum rule,  
behaves properly  as a decreasing function 
of $U$ at $U/(2 \pi t) \lesssim 0.9$. 
In this range of $U$, the results obtained with 
the integration of the DOS (not shown) agrees with the ones in the panel (b), 
and we can conclude that the second-order perturbation theory is applicable.  
The conductance shows the peak when a resonance peak passes 
through the Fermi level, and shows a minimum at half-filling, 
i.e., at $U=-2\epsilon_d$  or $\Delta N_{\rm tot}=32$.  
Due to the periodic boundary condition in the $y$-direction, 
we can see the contribution of each subband separately. 
In the case of $M=4$, 
the subband energy  $\epsilon_m = -2 t_y \cos (2\pi m/M)$ is given 
by  $\epsilon_{\rm I}=2t_y$, 
$\epsilon_{\rm II}=\epsilon_{\rm III}=0$, 
and $\epsilon_{\rm IV}=-2t_y$, where the labels are assigned as 
I ($m=2$), II ($m=1$), III ($m=-1$), and IV ($m=0$).
In Fig.\ \ref{fig:fig40} 
the results of the conductance are shown separately, 
where $\epsilon_d/t$ is taken to be (a) $-0.3$, (b) $-0.7$, and (c) $-1.0$. 
The resonance states belonging to the mode I ($\bullet$) 
and mode IV ($\triangle$) are sharp, 
while the peaks in the curve for the mode II and III ($\circ$)
are broad. 
Note that the two degenerate subbands, II and III,  
make the same contribution. 
Comparing the panels (a)--(c),
we see that the number of peaks increases 
when the onsite energy $\epsilon_d$ moves deep inside the Fermi level.
This is because the resonant states below the Fermi level 
can cross the Fermi level when the repulsive interaction $U$ increases.
In Fig.\ \ref{fig:fig30}, 
the local charge is plotted separately for each subband, 
where $\epsilon_d/t$ is taken to be ($\bullet$) 
$-0.3$,  ($\circ$) $-0.7$, and  ($\triangle$) $-1.0$. 
Naturally, the number of electrons in the subbands 
at low (high) energies is large (small). 
A step-like behavior is seen for the mode I and IV. 
This is because the occupation of the 
sharp resonance state changes rapidly from $2$ to $0$, 
when it crosses the Fermi level.

\begin{figure}[tb]
\begin{center}
\leavevmode
\includegraphics[width=0.6\linewidth, clip, 
trim = 1.5cm 14cm 1cm 1cm]{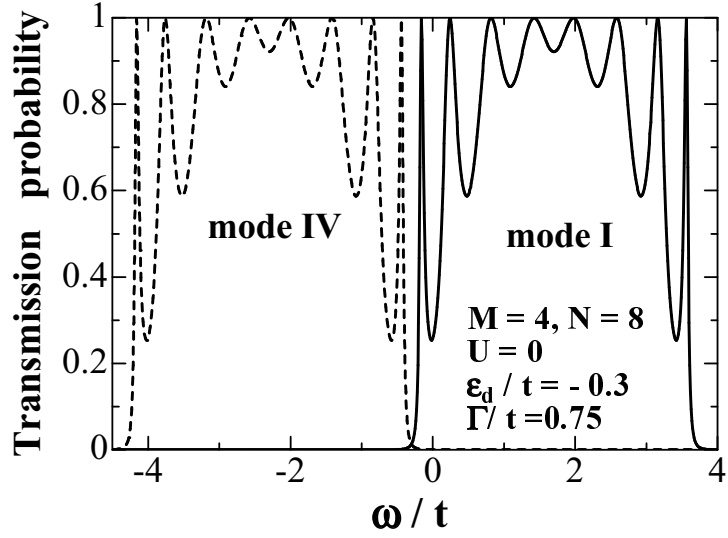}
\caption{Transmission probability 
vs $\omega$ for the noninteracting system of the size ($M=4$, $N=8$), 
 for $\epsilon_d/t=-0.3$. Here, only the contribution of 
the mode I (solid line)  and mode IV (dashed line) are plotted. 
The contribution of the modes II and III corresponds 
exactly to the dashed line in Fig.\ \ref{fig:fig19} that has plotted 
for a one-dimensional chain.
}
\label{fig:fig39}
\end{center}
\end{figure}

\begin{figure}[tb]
\begin{center}
\leavevmode
\includegraphics[width=0.7\linewidth, clip, 
trim = 1.5cm 6cm 0cm 1cm]{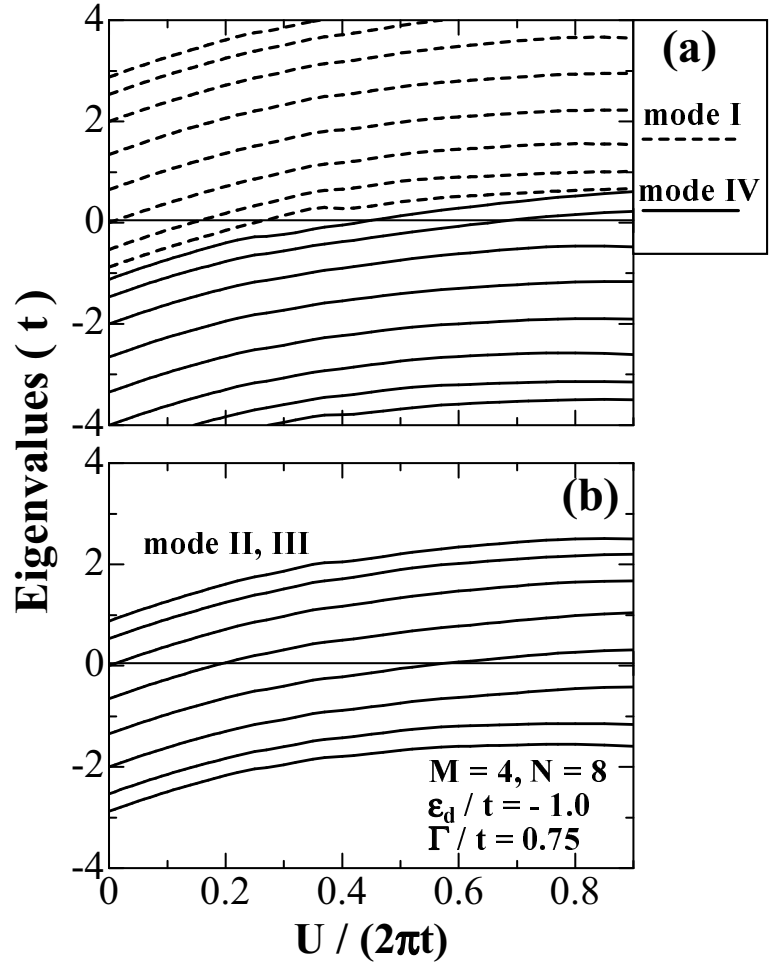}
\caption{Eigenvalues 
of $\widehat{\mbox{\boldmath ${\cal H}$}}_{\rm C}^{\rm eff}$ 
for the two-dimensional Hubbard model of the size ($M=4$, $N=8$).
Here, (a) the contribution of the modes I and IV,
 and (b) that of the degenerate modes II and III.
The onsite energy is taken to be $\epsilon_d / t$ = $-1.0$.
}
\label{fig:fig38}
\end{center}
\end{figure}

The feature of the resonance peaks can be compared with 
the transmission probability in the noninteracting case $U=0$
which is shown in Fig.\ \ref{fig:fig39} for $\epsilon_d /t = -0.3$. 
In this figure, only the contribution of 
the mode I (solid line) and mode IV (dashed line)
is plotted. The contribution of the mode II and III, 
which is not shown in Fig.\ \ref{fig:fig39}, 
is identical to the dashed line in Fig.\ \ref{fig:fig19} 
because of $\epsilon_{\rm II}=\epsilon_{\rm III}=0$.
The sharp peaks seen in the curves for the mode I and IV 
in Fig.\ \ref{fig:fig40} correspond to 
the ones at the band edge of the noninteracting subband-spectrum.
Furthermore,  
the broad peaks seen in Fig.\ \ref{fig:fig40} for the mode II and III 
correspond to the ones in the middle of 
the noninteracting subband-spectrum.
Therefore, when the interaction is switched on, 
the resonance peaks which already exist in the noninteracting system 
still remain changing their position and shape gradually. 
As discussed in the previous subsections, 
the eigenvalue of the effective 
Hamiltonian $\widehat{\mbox{\boldmath ${\cal H}$}}_{\rm C}^{\rm eff}$ 
defined by eq.\ (\ref{eq:effective}) has an additional information 
about the position of the resonance peaks.  
In Fig.\ \ref{fig:fig38}, the $U$ dependence of the eigenvalues 
is shown for $\epsilon_d / t=-1.0$. 
Among $N$ ($=8$) eigenvalues in each of the modes, 
those staying near the Fermi level contribute to the conductance. 
Comparing with the conductance shown in Fig.\ \ref{fig:fig40} (c), 
the resonant peaks seen in the contribution of the mode I (IV) 
corresponds to the lowest (highest) two 
eigenvalues which cross the Fermi level in Fig.\  \ref{fig:fig38} (a). 
Also, the broad peaks seen in the contribution of the mode II and III 
in Fig.\ \ref{fig:fig40} (c) correspond 
to the three intersections at Fermi level in Fig.\  \ref{fig:fig38} (b). 
Therefore,
from the eigenvalue just below the Fermi level, 
one can anticipate when the next peak will contribute 
to the tunneling current. 
This would not be possible 
if one sees only the conductance.

\begin{figure}[tb]
\begin{center}
\leavevmode
\includegraphics[width=0.6\linewidth, clip, 
trim = 1.5cm 3cm 1cm 1cm]{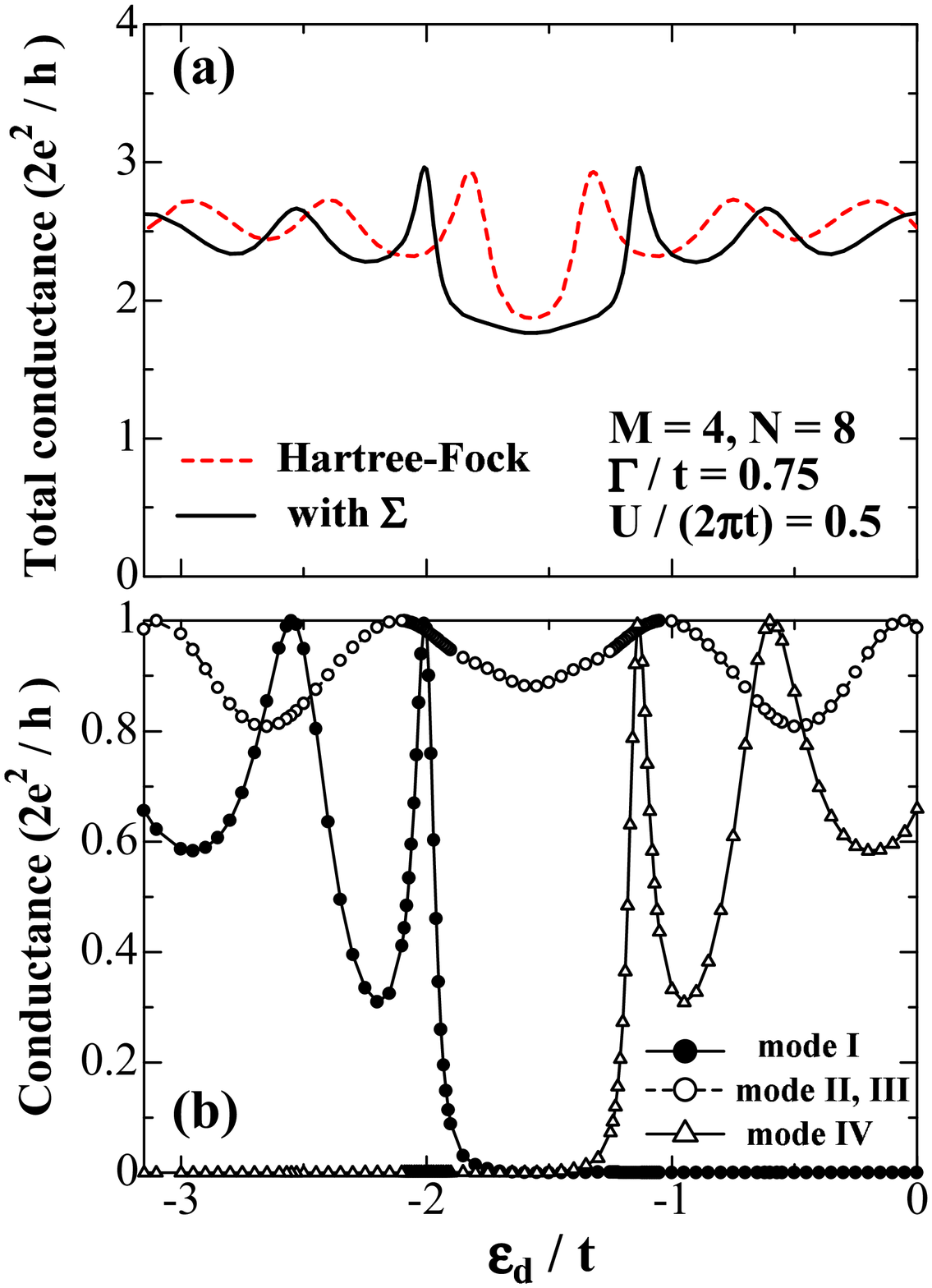}
\caption{
The $\epsilon_d$ dependence of 
the conductance through the two-dimensional cluster 
of the size ($M=4$, $N=8$). 
The upper panel (a) shows the total conductance, 
where the Hartree-Fock result (dashed line) 
is also plotted for comparison. 
The lower panel (b) shows the contribution of each of the modes.
The Coulomb interaction is taken to be  $U/(2 \pi t) = 0.5$,
so that the electron-hole symmetry holds 
at $\epsilon_d = -U/2\,$ ($\simeq -1.6\, t$). 
}
\label{fig:fig42}
\end{center}
\end{figure}

In Fig.\ \ref{fig:fig42} the conductance of  
the same two-dimensional lattice is shown as a function 
of the onsite energy $\epsilon_d$ taking the Coulomb 
interaction to be $U/(2 \pi t)=0.5$.
Note that, as mentioned in the previous subsection, 
 $\epsilon_d$ can be controlled by the gate voltage.
As seen in the lower panel (b), 
the contributions of the mode I and IV are symmetric 
with respect to the point $\epsilon_d = -U/2$ ($\simeq -1.6 \,t$) 
which corresponds to the half-filling. 
Around this point an {\em energy gap} of the order 
of $U$ opens up between these two subbands,
which is not present basically 
in the noninteracting case in Fig.\ \ref{fig:fig39}.
However, since the mode II and III 
carry the current even in this {\em energy gap} region,
the total conductance is finite in the plotted range 
of $\epsilon_d$ in the upper panel (a).
For comparison, 
the Hartree-Fock result of the total conductance
is also shown (dashed line) in Fig.\ \ref{fig:fig42} (a). 
Due to the self-energy correction beyond the Hartree-Fock approximation, 
the valley of the total conductance around the 
half-filling becomes deep and wide. 
Comparing with the one-dimensional chain 
of $N_{\rm}=1 \times 4$ [see Fig.\ \ref{fig:fig24} (b)], 
this tendency towards the development of 
a Mott-Hubbard insulator gap becomes clear 
in the two-dimensional lattice of $N_{\rm C} = 4 \times 8$.

\section{Summary}

In summary, 
we have studied the effects of electron correlation on the transport 
through a two-dimensional Hubbard model of finite size 
connected to two noninteracting leads, 
as a model for a quantum dot superlattice or materials on a nanometer scale.
Specifically, we have examined the ground-state 
properties away form half-filling by extending 
the previous work done at half-filling.
In the calculations, we determine first the unperturbed Green's function 
using the non-magnetic solution of 
the Hartree-Fock approximation,
and then take into account the second-order 
self-energy corrections with respect to the residual interaction 
that is not included in the mean field approximation.
This method works well for small $U$. However,   
as it is known for the single Anderson impurity, 
the radius of the convergence of the expansion with respect 
to $U$ decreases rapidly when the system 
goes away half-filling.~\cite{horva1} 
We have checked carefully the applicability of the method 
through the $U$ dependence of the local charge 
in the interacting region $\Delta N_{\rm tot}$, 
and study the transport property at small $U$ where 
$\Delta N_{\rm tot}$ behaves properly 
as a decreasing function of $U$.

We have calculated the zero-temperature conductance through the  
one- and two-dimensional Hubbard systems consisting 
of several tens of interacting sites. 
The conductance shows peaks caused by the resonant tunneling  
as a function of the onsite energy $\epsilon_d$
which can be controlled by the gate voltage, and 
as a function of the Coulomb interaction $U$.
Our numerical results show that 
the resonant peaks seen in the conductance 
have direct correspondence to 
the eigenvalues of the effective Hamiltonian 
which describes the quasi-particles of the local Fermi liquid.
Furthermore, from the eigenvalue we can get information 
about the resonance states inside the Fermi level.

The formulation used in the present study can be 
applied to various systems. 
We have also been studying how the randomness in the sample 
affects the transport through small interacting systems: 
real artificially systems seem more or less to be disordered. 
We have reported preliminary results obtained at half-filling,~\cite{YT2} 
and we are going to study the effects 
of the randomness away from half-filling.

\section*{Acknowledgements}
We would like to thank H.\ Ishii for valuable discussions. 
Numerical computation was partly performed 
at computation center of Nagoya University and 
at Yukawa Institute Computer Facility. 
This work was supported by the Grant-in-Aid 
for Scientific Research from the Ministry of Education, 
Science and Culture, Japan.

\clearpage

\clearpage

\clearpage

\clearpage


\begin{thebibliography}{99}



\bibitem{NL}
T. K. Ng and P. A. Lee: Phys.\ Rev.\ Lett.\ \textbf{61} (1988) 1768.

\bibitem{GR}
L. I. Glazman and M. E. Raikh: 
Pis'ma Zh.\  Eksp.\ Teor.\ Fiz.\ \textbf{47} (1988) 378
[JETP Lett.\ \textbf{47} (1988) 452].

\bibitem{Kawabata}
A. Kawabata: J.\ Phys.\ Soc.\ Jpn.\ \textbf{60} (1991) 3222.

\bibitem{MWL1-2}
Y. Meir, N. S. Wingreen and P. A. Lee: 
Phys.\ Rev.\ Lett.\ \textbf{66} (1991) 3048; 
\emph{ibid.\ } \textbf{70} (1993) 2601.

\bibitem{HDW2}
S. Hershfield, J. H. Davies and J. W. Wilkins: 
Phys.\ Rev.\ B \textbf{46} (1992) 7046.

\bibitem{Goldharber-Gordon}
D. Goldharber-Gordon, H. Shtrikman, D. Mahalu, D. Abusch-Magder,
U. Meirav and M. A. Kastner: Nature \textbf{391} (1998) 156.

\bibitem{Kouwenhoven}
S. M. Cronenwett, T. H. Oosterkamp and L. P. Kouwenhoven: 
Science \textbf{281} (1998) 540.

\bibitem{oosterkamp}
T. H. Oosterkamp, T. Fujisawa, W. G. van der Wiel, K. Ishibashi, 
R. V. Hijman, S. Tarucha and L. P. Kouwenhoven: 
Nature \textbf{395} (1998) 873.

\bibitem{tokura}
Y. Tokura, D. G. Austing and S. Tarucha: 
J. Phys. Condens. Matter \textbf{11} (1999) 6023.

\bibitem{tamura1-4} 
K. Shiraishi, H. Tamura and H. Takayanagi: 
Appl. \ Phys. \ Lett. \textbf{78} (2001) 3702; 
H. Tamura, K.Shiraishi, T. Kimura and H. Takayanagi: 
Phys. \ Rev. B \textbf{65} (2002) 85324. 

\bibitem{kimura}
T. Kimura, H. Tamura, K. Kuroki, K. Shiraishi, H. Takayanagi and R. Arita: 
Phys. Rev. B \textbf{66} (2002) 132508

\bibitem{Izumida1-3}
W. Izumida, O. Sakai and Y. Shimizu: 
J.\ Phys.\ Soc.\ Jpn.\ \textbf{66} (1997) 717; 
\emph{ibid.\/} \textbf{67} (1998) 2444.

\bibitem{Izumida4-5}
W. Izumida and O. Sakai: 
Phys.\ Rev.\ B \textbf{62} (2000) 10260; 
W. Izumida, O. Sakai and S. Suzuki: 
J.\ Phys.\ Soc.\ Jpn.\ \textbf{70} (2001) 1045.

\bibitem{Sakai}
O. Sakai, S. Suzuki, W. Izumida and A. Oguri: 
J.\ Phys.\ Soc.\ Jpn.\ \textbf{68} (1999) 1640.

\bibitem{ao6}
A. Oguri: Phys.\ Rev.\ B \textbf{56} (1997) 13422 
[Errata: \textbf{58} (1998) 1690].  

\bibitem{Hewson}
A.\ C.\ Hewson: 
{\em  The Kondo Problem to Heavy Fermions\/} 
(Cambridge University Press, Cambridge, 1993). 

\bibitem{YamadaYosida}
K. Yamada: Prog.\ Theor.\ Phys.\  {\bf 53} (1975) 970; 
Prog.\ Theor.\ Phys.\  {\bf 54} (1975) 316; 
K. Yosida and K. Yamada: 
Prog.\ Theor.\ Phys.\ {\bf 53} (1975) 1286. 

\bibitem{ZlaticHorvatic}
V. Zlati\'c and B. Horvati\'c: 
Phys.\ Rev.\ B {\bf 28} (1983) 6904.

\bibitem{ao11-ao12}
A.\ Oguri: Phys.\ Rev.\ B {\bf 64} (2001) 153305; 
J.\ Phys.\ Soc.\ Jpn.\ \textbf{71} (2002) 2969.

\bibitem{Eliashberg}
G. M. \'{E}liashberg: Zh.\  Eksp.\ Teor.\ Fiz.\ {\bf 41} (1961) 1241 
[JETP {\bf 14} (1962) 886].

\bibitem{ao10}
A. Oguri: 
J.\ Phys.\ Soc.\ Jpn.\ \textbf{70} (2001) 2666.

\bibitem{YT}
Y. Tanaka, A. Oguri and H. Ishii: 
J.\ Phys.\ Soc.\ Jpn.\ \textbf{71} (2002) 211.

\bibitem{ao7-ao9}
A. Oguri: Phys.\ Rev.\ B \textbf{59} (1999) 12240; 
Phys.\ Rev.\ B \textbf{63} (2001) 115305 
[Errata: \textbf{63} (2001) 249901]. 

\bibitem{HohenbergKohn}
P. Hohenberg and W. Kohn: 
Phys.\ Rev.\ {\bf 136} (1964) B864.

\bibitem{horva1}
B. Horvati\'c and V. Zlati\'c: 
\ Phys. \ Status \ Solidi \ B \textbf{99} (1980) 251.

\bibitem{Yeyati}
A. Levy Yeyati, A. Mart\'in and F. Flores: 
Phys.\ Rev.\ Lett.\ \textbf{71} (1993) 2991.

\bibitem{Saso}
O. Takagi and T. Saso: J.\ Phys.\ Soc.\ Jpn.\ \textbf{68} (1999) 1997.

\bibitem{LangerAmbegaokar}
J. S. Langer and V. Ambegaokar: 
Phys.\ Rev.\ {\bf 121} (1961) 1090.

\bibitem{Anderson}
P. W. Anderson: Phys.\ Rev.\ {\bf 124} (1961) 41.

\bibitem{YT2}
Y. Tanaka and  A. Oguri: Physica E {\bf 18} (2003) 300.  



\end{thebibliography}
\end{document}